\documentclass[aps,prx,reprint,superscriptaddress,nofootinbib,twocolumn]{revtex4-2}
\usepackage{blindtext}
\usepackage{lipsum}
\usepackage{graphics}
\usepackage{amsmath}
\usepackage{graphicx}
\usepackage{graphics}
\usepackage{amssymb}
\usepackage{verbatim}
\usepackage{framed}
\usepackage{float}
\usepackage{braket}
\usepackage{bm}
\usepackage{mathtools}
\usepackage{bbm}
\usepackage{multirow}
\usepackage[toc,page]{appendix}
\usepackage{CJKutf8}

\usepackage[colorlinks=true,citecolor=dodgerblue,linkcolor=dodgerblue,urlcolor=dodgerblue,pdftitle={TriHubHof}]{hyperref}

\makeatletter
\def\l@subsubsection#1#2{} 
\makeatother

\usepackage[dvipsnames]{xcolor}
\usepackage[normalem]{ulem}
\usepackage{wasysym} 
\usepackage[export]{adjustbox}

\makeatletter
\DeclareFontFamily{OMX}{MnSymbolE}{}
\DeclareSymbolFont{MnLargeSymbols}{OMX}{MnSymbolE}{m}{n}
\SetSymbolFont{MnLargeSymbols}{bold}{OMX}{MnSymbolE}{b}{n}
\DeclareFontShape{OMX}{MnSymbolE}{m}{n}{
    <-6>  MnSymbolE5
   <6-7>  MnSymbolE6
   <7-8>  MnSymbolE7
   <8-9>  MnSymbolE8
   <9-10> MnSymbolE9
  <10-12> MnSymbolE10
  <12->   MnSymbolE12
}{}
\DeclareFontShape{OMX}{MnSymbolE}{b}{n}{
    <-6>  MnSymbolE-Bold5
   <6-7>  MnSymbolE-Bold6
   <7-8>  MnSymbolE-Bold7
   <8-9>  MnSymbolE-Bold8
   <9-10> MnSymbolE-Bold9
  <10-12> MnSymbolE-Bold10
  <12->   MnSymbolE-Bold12
}{}

\let\llangle\@undefined
\let\rrangle\@undefined
\DeclareMathDelimiter{\llangle}{\mathopen}%
                     {MnLargeSymbols}{'164}{MnLargeSymbols}{'164}
\DeclareMathDelimiter{\rrangle}{\mathclose}%
                     {MnLargeSymbols}{'171}{MnLargeSymbols}{'171}
\makeatother

\usepackage{ulem}

\usepackage{amssymb}
\usepackage{pifont}

\newcommand{\bmr}{\bm{r}}
\newcommand{\bma}{\bm{a}}
\newcommand{\bmb}{\bm{b}}

\newcommand{\mcO}{\mathcal{O}}
\newcommand{\mcP}{\mathcal{P}}

\newcommand{\R}{\mathbb{R}}
\newcommand{\Z}{\mathbb{Z}}

\newcommand{\n}[1]{\left| #1 \right|}

\definecolor{orange(ryb)}{HTML}{FFA500}
\definecolor{dodgerblue}{HTML}{1E90FF}
\definecolor{pinkerton}{HTML}{EC368D}
\definecolor{forest}{HTML}{6DD189}

\definecolor{edit}{HTML}{000000}

\usepackage{tikz}
\usepackage{tikz-cd}
\usetikzlibrary{arrows}
\usetikzlibrary{snakes}
\usetikzlibrary{intersections}
\usetikzlibrary{shapes.geometric}
\usetikzlibrary{decorations.pathmorphing, patterns,shapes}
\usetikzlibrary{decorations.markings}

\tikzset{
	partial ellipse/.style args={#1:#2:#3}{
		insert path={+ (#1:#3) arc (#1:#2:#3)}
	}
}

\tikzset{
	mid arrow/.style={postaction={decorate,decoration={
				markings,
				mark=at position .575 with {\arrow[#1]{stealth}}
	}}},
	near arrow/.style={postaction={decorate,decoration={
				markings,
				mark=at position .275 with {\arrow[#1]{stealth}}
	}}},
	far arrow/.style={postaction={decorate,decoration={
				markings,
				mark=at position .800 with {\arrow[#1]{stealth}}
	}}},
}

\pgfdeclarepatternformonly{south west lines}{\pgfqpoint{-0pt}{-0pt}}{\pgfqpoint{3pt}{3pt}}{\pgfqpoint{3pt}{3pt}}{
	\pgfsetlinewidth{0.4pt}
	\pgfpathmoveto{\pgfqpoint{0pt}{0pt}}
	\pgfpathlineto{\pgfqpoint{3pt}{3pt}}
	\pgfpathmoveto{\pgfqpoint{2.8pt}{-.2pt}}
	\pgfpathlineto{\pgfqpoint{3.2pt}{.2pt}}
	\pgfpathmoveto{\pgfqpoint{-.2pt}{2.8pt}}
	\pgfpathlineto{\pgfqpoint{.2pt}{3.2pt}}
	\pgfusepath{stroke}}

\makeatletter
\renewcommand\onecolumngrid{
\do@columngrid{one}{\@ne}%
\def\set@footnotewidth{\onecolumngrid}
\def\footnoterule{\kern-6pt\hrule width 1.5in\kern6pt}%
}

\renewcommand\twocolumngrid{
        \def\footnoterule{
        \dimen@\skip\footins\divide\dimen@\thr@@
        \kern-\dimen@\hrule width.5in\kern\dimen@}
        \do@columngrid{mlt}{\tw@}
}
\makeatother

\begin{document}

\title{Chiral Spin Liquid and Quantum Phase Transition in the Triangular Lattice Hofstadter-Hubbard Model}

\author{Stefan Divic}
\affiliation{Department of Physics, University of California, Berkeley, CA 94720, USA}
\affiliation{Center for Computational Quantum Physics, Flatiron Institute, New York, New York 10010, USA}
\author{Tomohiro Soejima (\begin{CJK*}{UTF8}{bsmi}副島智大\end{CJK*})}
\affiliation{Department of Physics, Harvard University, Cambridge, MA 02138, USA}
\author{Valentin Cr\'epel}
\affiliation{Center for Computational Quantum Physics, Flatiron Institute, New York, New York 10010, USA}
\author{Michael P. Zaletel}
\affiliation{Department of Physics, University of California, Berkeley, CA 94720, USA}
\affiliation{Material Science Division, Lawrence Berkeley National Laboratory, Berkeley, CA 94720, USA}
\author{Andrew Millis}
\affiliation{Center for Computational Quantum Physics, Flatiron Institute, New York, New York 10010, USA}
\affiliation{Department of Physics, Columbia University, New York, NY 10027, USA}

\date{\today}

\begin{abstract}
Recent advances in moir\'e engineering motivate the study of lattice models of strongly-correlated electrons subjected to substantial orbital magnetic flux. We analyze the triangular lattice Hofstadter-Hubbard model at one-quarter flux quantum per plaquette and a density of one electron per site, where a chiral spin liquid phase may exist between weak-coupling integer quantum Hall and strong-coupling 120$^\circ$ antiferromagnetic phases.
We use matrix product state methods and analytical arguments to investigate this model compactified to cylinders of finite circumference.
We uncover a glide particle-hole symmetry operation which, we argue, is spontaneously broken at the quantum Hall to spin liquid transition on odd-circumference cylinders. We numerically verify the spontaneous symmetry breaking and further demonstrate that this transition is associated with algebraic long-range correlations of various spin-singlet, charge-neutral operators.
For even-circumference cylinders, the transition becomes a crossover associated with a large correlation length that grows substantially with circumference.
Our findings suggest that in the two-dimensional limit, the transition to a chiral spin liquid phase is continuous and features critical fluctuations of the current.
\end{abstract}

\maketitle

\textit{Introduction---}The Kalmeyer-Laughlin chiral spin liquid (CSL) is one of the earliest and most important examples of a topologically-ordered quantum state~\cite{Kalmeyer1987, Laughlin1989, Wen1989} and has been extensively studied in the context of spin models~\cite{Yang1993, WenBook2007, motrunich2006spinliquid, He2014}.
Recently, considerable effort has been devoted to identifying electronic settings in which the CSL may emerge~\cite{Szasz2020, Szasz2021, zou1990charged, bauer2014chiral, zhu2024chiral, hickey2016haldane, grushin2023amorphous,Zang21,crepel2024spinon,Ghiotto_2021Nature,Zhang2021,Kuhlenkamp2024}.
These studies set the stage for the investigation of charge fluctuation-driven transitions out of the CSL phase into other phases such as metals~\cite{Szasz2020}, superconductors~\cite{song2021dopingCSL}, exotic charge-density waves~\cite{SongZhang2023}, and quantum Hall states~\cite{He2011,Kuhlenkamp2024}.

\begin{figure}
\centering
\includegraphics[width=\columnwidth]{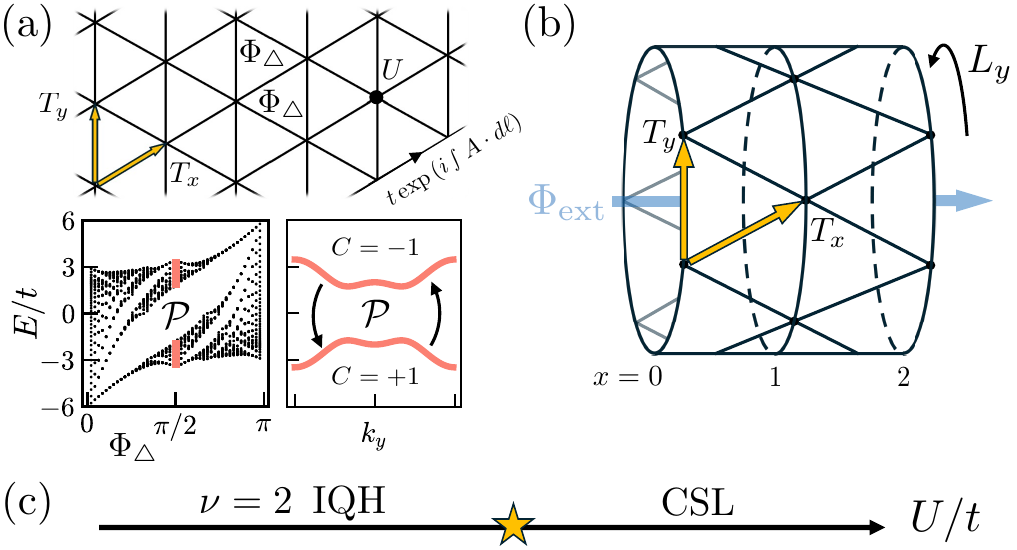}
\caption{(a) Upper panel: The triangular Hofstadter-Hubbard model, with magnetic translations $T_{x,y}$, flux per triangle $\Phi_\triangle$, and Hubbard interaction $U$ indicated. Lower left panel: ``butterfly'' representation of non-interacting spectrum in the plane of $\Phi_\triangle$ and energy $E$. The bands at $\Phi_\triangle = \pi/2$  enjoy a doubled ``magnetic'' unit cell and particle-hole symmetry $\mathcal{P}$ relating opposite $C=\pm 1$ bands [lower right panel].
(b) Finite segment of the circumference-$L_y$ cylinder threaded by $\Phi_\text{ext}$ external flux, where $T_y$ translates around the cylinder and $x\in\mathbb{Z}$ indexes rings.
(c) 2+1D phase diagram at a density of $n=1$ electron per site as a function of $U/t$ showing the integer quantum Hall (IQH) and chiral spin liquid (CSL) phases, with their transition marked by a star.
}
\label{fig:MainMessage}
\end{figure}

The triangular Hofstadter-Hubbard model is thought to host both conventional and CSL electronic phases in the regime of strong orbital magnetic flux~\cite{Cookmeyer2021CSL,sen1995chiral,motrunich2006spinliquid, WietekHeisenbergChiral2017,Gong2017,Saadatmand2017,Huang2024}.
It is parameterized by an on-site interaction $U$ and hoppings of magnitude $t$ and phases consistent with a magnetic flux of $\Phi_\triangle$ per triangular plaquette (see Fig.~\ref{fig:MainMessage}a). When $\Phi_\triangle=\pi/2$, this system realizes a spin-singlet integer quantum Hall (IQH) insulator at small $U$ and density of one electron per site, with $\sigma_{xy}=2e^2/h$.
At very large $U$, the state is a non-topological Mott insulator with a large charge gap, $120^\circ$ antiferromagnetic (AF) order, and gapless spin excitations described by a nearest-neighbor Heisenberg exchange of magnitude $\sim t^2/U$~\cite{WietekHeisenbergChiral2017,Gong2017,Saadatmand2017,Huang2024}.

\begin{figure*}
\centering
\includegraphics[width=\textwidth]{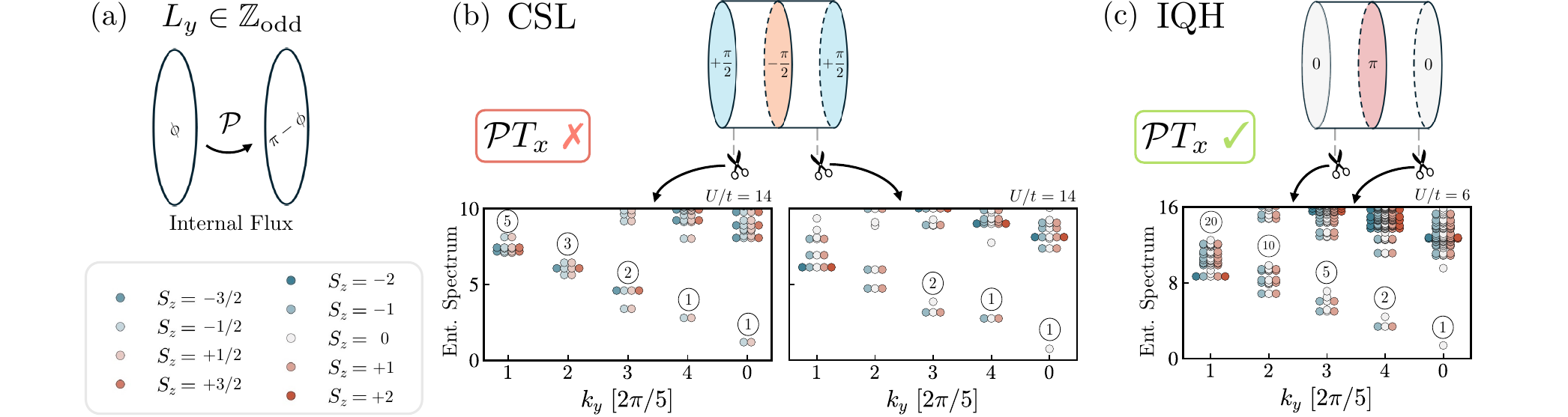}
\caption{Symmetry breaking and entanglement spectrum (ES) on odd-$L_y$ cylinders with external ring fluxes alternating between $0$ and $\pi$. (a) The particle-hole operation $\mathcal{P}$ transforms the internal gauge flux through an odd-$L_y$ cylinder ring from $\phi$ to $\pi - \phi$.
(b) Upper panel: In an odd-$L_y$ CSL ground state, the saddle-point pattern $(\pi/2,-\pi/2,\dots)$ of these internal fluxes at strong coupling spontaneously breaks $\mathcal{P} T_x$ symmetry.
Lower panel: the low-lying entanglement eigenvalues on the $L_y=5$ cylinder at spatial cuts between adjacent pairs of rings, which not only exhibit the approximate degeneracy expected for the CSL at each fixed $S_z$ [circled numerals], but also signal spontaneous breaking of $\mathcal{P} T_x$.
(c) The internal flux configuration [upper panel] and low-lying entanglement eigenvalues [lower panel] of the IQH at $L_y=5$. The ES at the two spatial cuts are equivalent.
}
\label{fig:ES}
\end{figure*}

A qualitative argument for the existence of an intermediate CSL phase is that the leading correction to the large-$U$ effective Heisenberg spin Hamiltonian is a chiral term $S \cdot S \times S$ with coefficient $J_\triangle\sim t^3 \sin\Phi_\triangle /U^2$, which explicitly breaks both time-reversal and parity symmetry and favors a CSL state~\cite{sen1995chiral, motrunich2006spinliquid, Hu2016, WietekHeisenbergChiral2017, Gong2017, Saadatmand2017, Huang2024}. While a CSL has also been reported in the time reversal-symmetric $\Phi_\triangle=0$ model at intermediate coupling between a Fermi liquid and $120^\circ$ AF~\cite{Szasz2020,Chen2022,Kadow2022}, where it is putatively stabilized by four-spin interactions~\cite{Cookmeyer2021CSL}, other magnetic and spin liquid phases are so close in energy to the CSL~\cite{Gong2017,WietekHeisenbergChiral2017,SaadatmandMcCulloch2017} that its existence at $\Phi_\triangle=0$ remains debated~\cite{Shirakawa2017,Wietek2021}. In contrast, Kuhlenkamp \textit{et al.} very recently presented cylinder iDMRG evidence of a robust intermediate CSL phase starting from $U\gtrsim 8.5t$ at flux $\Phi_\triangle = \pi/3$ and $U\gtrsim 10.5t$ at $\Phi_\triangle = \pi/2$~\cite{Kuhlenkamp2024}.

The nature of the IQH-CSL phase transition at $\Phi_\triangle=\pi/2$, however, is not well understood. Parton mean field theories have been proposed for related square lattice and honeycomb systems~\cite{He2011,chen2016synthetic}. Beyond this, there has been very little direct investigation into the nature of the putative critical point from either theory or numerics. In particular, the iDMRG investigation of Ref.~\cite{Kuhlenkamp2024} unearthed the IQH-CSL transition only as a crossover indicated by a weak maximum in a correlation length.

In this work, we use a compactification to infinite-length, finite-circumference cylinders (see Fig.~\ref{fig:MainMessage}b) in combination with a symmetry analysis and large-scale iDMRG simulations~\cite{White1992,White1993,mcculloch2008} to characterize the IQH and CSL states, as well as the transition between them.
We show that a combination of particle-hole interchange and translation is spontaneously broken on odd-circumference cylinders in the CSL phase and present evidence that the IQH-CSL transition is in fact continuous in these geometries, with no intervening phase. We argue that the critical fluctuations are spin-singlet, with both spin and electron correlation functions decaying exponentially with distance. Considering also even-circumference cylinders, we argue that in 2+1D the IQH and CSL phases are separated by a continuous quantum phase transition that includes critical current fluctuations.

\begin{figure*}
\centering
\includegraphics[width=\textwidth]{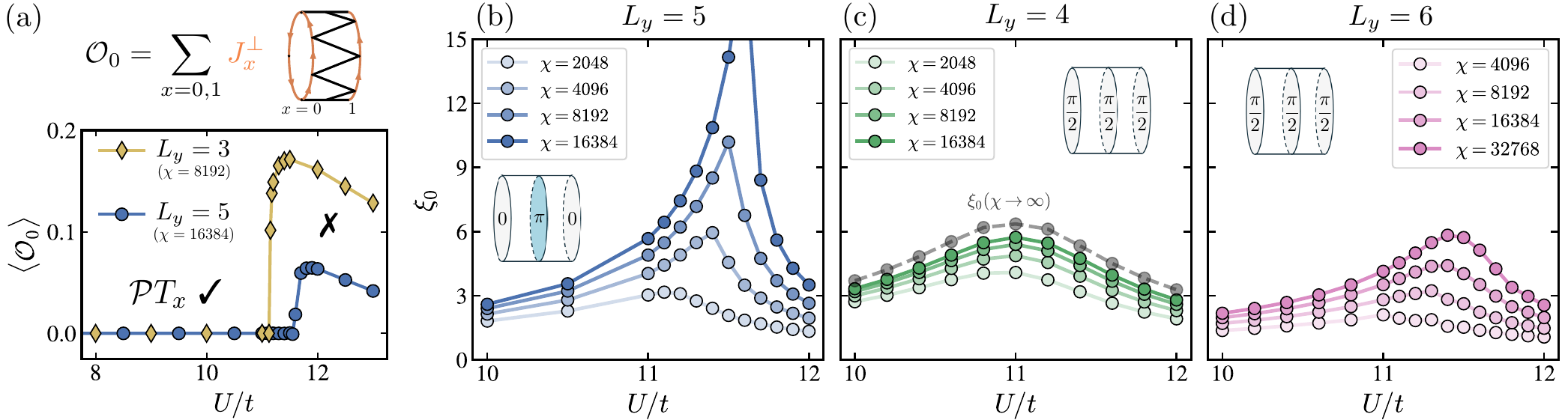}
\caption{Evidence for the IQH-CSL continuous phase transition. (a) Upper panel: Sketch of two cylinder rings, illustrating the $\mathbb{Z}_2$ order parameter $\mcO_0$ defined as the total circumferential current operator [orange arrows]. Lower panel: $\langle \mcO_0\rangle$ versus $U/t$ for the odd-circumference cylinders, with $\langle \mcO_0\rangle\neq0$ implying $\mcP T_x$ is spontaneously broken.
(b-d) MPS transfer matrix correlation length in the $(Q,S_z)=0$ charge sector for various MPS bond dimensions [increasing light to dark] and extrapolated to infinite bond dimension for $L_y=4$ [gray points]; the insets specify the external magnetic flux through the cylinder rings.
}
\label{fig:OP_and_xi}
\end{figure*}

\smallskip
\textit{Model, particle-hole symmetry and compactification---}We study the Hofstadter-Hubbard model on the triangular lattice, defined pictorially in Fig.~\ref{fig:MainMessage}(a). In this model, electrons hop between nearest-neighboring sites and experience an Aharonov-Bohm phase of $\exp(i\Phi_\triangle)$ upon orbiting a triangular plaquette (see Supplemental Material Sec.~A for details~\cite{SoM}). We focus on the consequences of the orbital motion, setting the Zeeman coupling to zero~\cite{Zhang2021,Kuhlenkamp2024}. We fix $\Phi_\triangle=\pi/2$~\footnote{For a moir\'e lattice constant $a$, the magnetic field required for $\Phi_\triangle = \pi/2$ flux is $B = 4a^{-2}/\sqrt{3} \cdot h/4e$. For example, $B=15$~T requires $a=12.6$~nm, achievable in twisted transition metal dichalcogenides~\cite{2024SciFoutty}. See Refs.~\cite{Zhang2021,Kuhlenkamp2024} for discussion of potential material realizations.}, which endows the Hamiltonian with a unitary particle-hole (PH) symmetry $\mcP$ chosen to act as
\begin{align} \label{eq:particlehole}
    \mcP c^\dagger_{x,y} \mcP^{-1} = (-1)^{x+y} (i\sigma_y) c_{x,y}, \quad \mcP i \mcP^{-1} = +i,
\end{align}
(the spin rotation $i\sigma_y$ is included in $\mcP$ so it leaves the spin operator $\bm S = \frac{1}{2}c^\dagger \bm \sigma c$ invariant; see Supplemental Material Sec.~C~\cite{SoM}). Denoting translation along the two non-parallel directions shown in Fig.~\ref{fig:MainMessage} as $T_{x,y}$, which we represent in a gauge where
\begin{align}
    T_x c^\dag_{x,y} T_x^{-1} = (-1)^y c^\dag_{x+1,y},\quad T_y c^\dag_{x,y} T_y^{-1} = c^\dag_{x,y+1},
\end{align}
we obtain the required magnetic translation algebra $T_x T_y = (-1)^{N_F} T_y T_x$, where $N_F$ is the fermion number.
The commuting translations $(T_x)^2, T_y$ define a larger ``magnetic'' unit cell, yielding two sub-bands with Chern numbers $C=\pm 1$ related by $\mcP$ (Fig.~\ref{fig:MainMessage}a). We fix the filling to $n=1$ electron per lattice site, corresponding to full filling of the lowest spin-degenerate magnetic sub-band.

Dimensional reduction can break the $T_x$ symmetry. Consider the closed ``tin can'' shaped surface demarcated by the pair of adjacent lattice rings at $x$ and $x+1$. Applying Gauss's law on this surface~\cite{Lu2020, SahayDivic2023}, with $\Phi_x$ denoting the external flux through the cylinder ring at $x$, we have $\Phi_{x+1} + 2 L_y\Phi_\Delta \equiv \Phi_x$ modulo $2\pi$. The ring fluxes are therefore uniform when $L_y$ is even, but staggered by $\pi$ when $L_y$ is odd. In the latter case, $T_x$ is explicitly broken.

Consider the odd circumference case; henceforth, we fix the external ring fluxes to alternate between $0$ and $\pi$. Though $T_x$ is broken, the combined translation-like operation $\mcP T_x$ (``glide-PH'') commutes with the Hamiltonian (see Supplemental Material Sec.~C~\cite{SoM}).
The IQH state, being adiabatically connected to the unique ground state of the non-interacting Hamiltonian, respects $\mcP T_x$. However, the odd-$L_y$ CSL phase breaks $\mcP T_x$ spontaneously. To see this, note that in the CSL phase the charge fluctuations are gapped; when $U$ is large, the low energy degrees of freedom are described by an effective spin model~\cite{motrunich2006spinliquid}.
Since $\mcP \bm{S} \mcP^{-1} = \bm{S}$, this spin model has a pure translation symmetry $t_x$.
As each ring contains an odd number of spin-$1/2$ moments, the Lieb-Schulz-Mattis theorem requires that any gapped ground state be at least twofold degenerate~\cite{Lieb1961}, \textit{e.g.}, by breaking $t_x$ spontaneously. The same conclusion can be drawn from the anyon content of the CSL: the presence of one semion per unit cell~\cite{XGW2002,XGW2003,Essin2013,Lu2020} causes $t_x$ to permute the two minimally-entangled ground states~\cite{zhang2012quasiparticle,Zaletel2015,Zaletel2017}.

We can also interpret the CSL symmetry breaking in terms of the parton decomposition of the electron into a bosonic chargon $b$ carrying the charge and a fermionic spinon $f$ carrying the spin, $c_\sigma = b f_\sigma$~\cite{Florens2004,lee2005u,Senthil2008}. This requires introducing an \textit{internal} gauge field $a$. In both the CSL and IQH phases, the saddle point configuration of the internal gauge field yields an internal flux through the triangular surface plaquettes that matches the external flux, $\phi_\triangle = \pi/2$~\cite{WenBook2007,motrunich2006spinliquid,song2021dopingCSL}. Moreover, assuming $L_y$ is odd, the internal flux through adjacent cylinder \textit{rings} may be written as $(\phi, \phi+\pi)$, again differing by $\pi$ due to Gauss's law.
Since the CSL ground states are adiabatically connected to the Mott limit where $\mcP$ must act as the identity operator, at strong coupling their internal ring fluxes will energetically lock to $(\pi/2,-\pi/2)$ or $(-\pi/2,\pi/2)$, which are both invariant under the action of $\mcP$, namely $\phi\to \pi-\phi$ (Fig.~\ref{fig:ES}a).
These two configurations are related by $\mcP T_x \simeq t_x$, indicating spontaneous symmetry breaking (SSB) in these odd-$L_y$ geometries (Fig.~\ref{fig:ES}b).
In the IQH phase where the chargon is condensed, the Higgs mechanism instead pins together the internal and external gauge fields~\cite{Anderson1963,Higgs1964,sachdev_quantum_2023}. The resulting internal ring flux pattern $(0, \pi)$ respects $\mathcal{P} T_x$, consistent with a unique, gapped ground state (Fig.~\ref{fig:ES}c, top).

\smallskip
\textit{Numerical observation of phase transition---}We have used the iDMRG algorithm to determine the ground state on cylinders of circumference $L_y=3,4,5,6$. For the numerically studied range $U \leq 14$, we find no sign of the strong-coupling $120^\circ$ AF, so we focus on the IQH and CSL phases. We study the entanglement spectrum (ES)~\cite{Calabrese2008,LiHaldane2008} at a spatial bipartition of the system into left and right semi-infinite cylinders~\cite{Szasz2020, Kuhlenkamp2024}.
The low-lying part of the ES corresponds to the conformal field theory (CFT) of the physical edge and exhibits degeneracies characteristic of the edge CFT~\cite{KitaevPreskill_TEE,LiHaldane2008,Fidkowski2010,Chandran2011,Qi2012,Sterdyniak2012,SwingleSenthil2012,Tu2013}.

Consistent with the above expectations, the CSL ground state spontaneously breaks $\mcP T_x$ symmetry on odd-$L_y$ cylinders: the low-lying entanglement eigenstates carry half-integer (integer) spin on bipartitions made to the right of even (odd) cylinder rings $x$ (see Supplemental Material Sec.~D~\cite{SoM}). At the even cuts, the number of nearly-degenerate states with fixed $S_z=\hbar/2$ are $(1,1,2,3,5,\dots)$ at consecutive momenta; the counting of $S_z=0$ states on the odd cuts is similarly $(1,1,2,\dots)$ (see Fig.~\ref{fig:ES}b, bottom). This pattern is characteristic of the free boson CFT~\cite{XGW1991}. In contrast, the low-lying ES of the IQH features states with both integer and half-integer spin (see Supplemental Material Sec.~D~\cite{SoM}). Consistent with unbroken $\mcP T_x$ symmetry, the spectrum is equivalent on every bipartition (Fig.~\ref{fig:ES}c, bottom). The approximate degeneracies at fixed $S_z$ are $(1,2,5,10,20\dots)$, corresponding to two copies of the free boson CFT~\cite{crepel2018matrix,WuSreejithJain_2012_edge,BalatskyStone1991,Wen1992_IJMPB_edge}.

We also construct a $\mathbb{Z}_2$ order parameter for $\mcP T_x$, $\mathcal{O} = \sum_{x\in\mathbb{Z}} J^\perp_x$ where $J^\perp_x$ is the average current on the circumferential bonds at $x$.  
In the chosen gauge at $\Phi_\mathrm{ext}=0$:
\begin{align}
\label{eq:orderparameter}
    J^\perp_x = (-1)^x L^{-1}_y \sum_y i (c_{x,y}^{\dagger}c_{x,y+1} - \mathrm{h.c.}).
\end{align}
Since $(\mathcal{P}T_{x}) J^\perp_x (\mathcal{P}T_{x})^{\dagger} = -J^\perp_{x+1}$, then $\mathcal{O}$ is charged under $\mathcal{P}T_{x}$. Since the ground states are invariant under $(\mathcal{P}T_{x})^2$ so that $\langle J^\perp_{x}\rangle = \langle J^\perp_{x+2}\rangle$, it suffices to examine the condensation of $\mathcal{O}_0 = J^\perp_0 + J^\perp_1$. In the CSL phase, $\langle\mcO_{0}\rangle\neq 0$ (see Fig.~\ref{fig:OP_and_xi}a), signaling SSB at $L_y=3,5$.
Near the transition at odd $L_y$, the transfer matrix spectrum~\cite{Zauner2015} contains a \textit{diverging} correlation length in the $(Q,S_z,k_y)=0$ charge sector (see Fig.~\ref{fig:OP_and_xi}(b) for $L_y=5$ and Supplemental Material Sec.~G for $L_y=3$~\cite{SoM}) consistent with the charge of the order parameter $\mcO$. Threading $\Phi_\text{ext}=\pi/2$ external flux through the odd-$L_y$ cylinder reduces the transition to a crossover since glide-PH symmetry is explicitly broken when $\Phi_\text{ext} \not\in \pi\mathbb{Z}$.

We further measure $\mcO_{\delta\mathrm{Heis}} = \sum_{m\in\mathbb{Z}} \mcO^{\delta\mathrm{Heis}}_{2m}$, defined by
\begin{align} \label{eq:stagHeis}
    \mcO^{\delta \text{Heis}}_x = L_y^{-1}\sum_{y} \bm{S}_{x,y}\cdot\bm{S}_{x+1,y}-\bm{S}_{x+1,y}\cdot\bm{S}_{x+2,y}.
\end{align}
The order parameter $\mcO_{\delta\mathrm{Heis}}$ is likewise charged under $\mcP T_x$ and condenses in the CSL phase for odd $L_y$ and $\Phi_\text{ext}=0$ (see Supplemental Material Sec.~E~\cite{SoM}). The two order parameters share the same symmetries and would be linearly coupled in a Ginzburg-Landau theory. We prefer $\mathcal{O}$ defined in Eq.~\eqref{eq:orderparameter} because it is larger in magnitude and simpler to express.

On the even circumference geometries $L_y = 4,6$ we reproduce the weak crossover, with a correlation length of $\lesssim 1.5$ lattice sites, found by Kuhlenkamp \textit{et al.} for $L_y = 6$ and $\Phi_\text{ext}=0$~\cite{Kuhlenkamp2024} (see Supplemental Material Sec.~F~\cite{SoM}).
Remarkably, upon threading $\Phi_\text{ext}=\pi/2$, the correlation length in the range $10 < U < 12$ increases dramatically, notably in the same $(Q,S_z)=0$ sector in which our odd-circumference correlation lengths diverge [see Fig.~\ref{fig:OP_and_xi}(c,d)]. At $L_y=4$, extrapolation to infinite bond dimension following Ref.~\cite{Rams2018} predicts a maximal extrapolated value of $\xi_{L_y=4}^{\chi\to\infty} = 6.6$ in units of cylinder rings (see Supplemental Material Sec.~F~\cite{SoM}). While $\xi = 5.8$ at $L_y=6$ at the largest bond dimension $\chi=32768$, the correlation length is still growing rapidly with $\chi$. This hampers the precise extrapolation of $\xi_{L_y=6}^{\chi\to\infty}$ but also indicates it substantially exceeds $\xi_{L_y=4}^{\chi\to\infty}$.
This suggests that the larger even-$L_y$ geometries will exhibit a sequence of increasingly-sharp crossovers toward 2+1D criticality.

\begin{figure}
\centering
\includegraphics[width=\columnwidth]{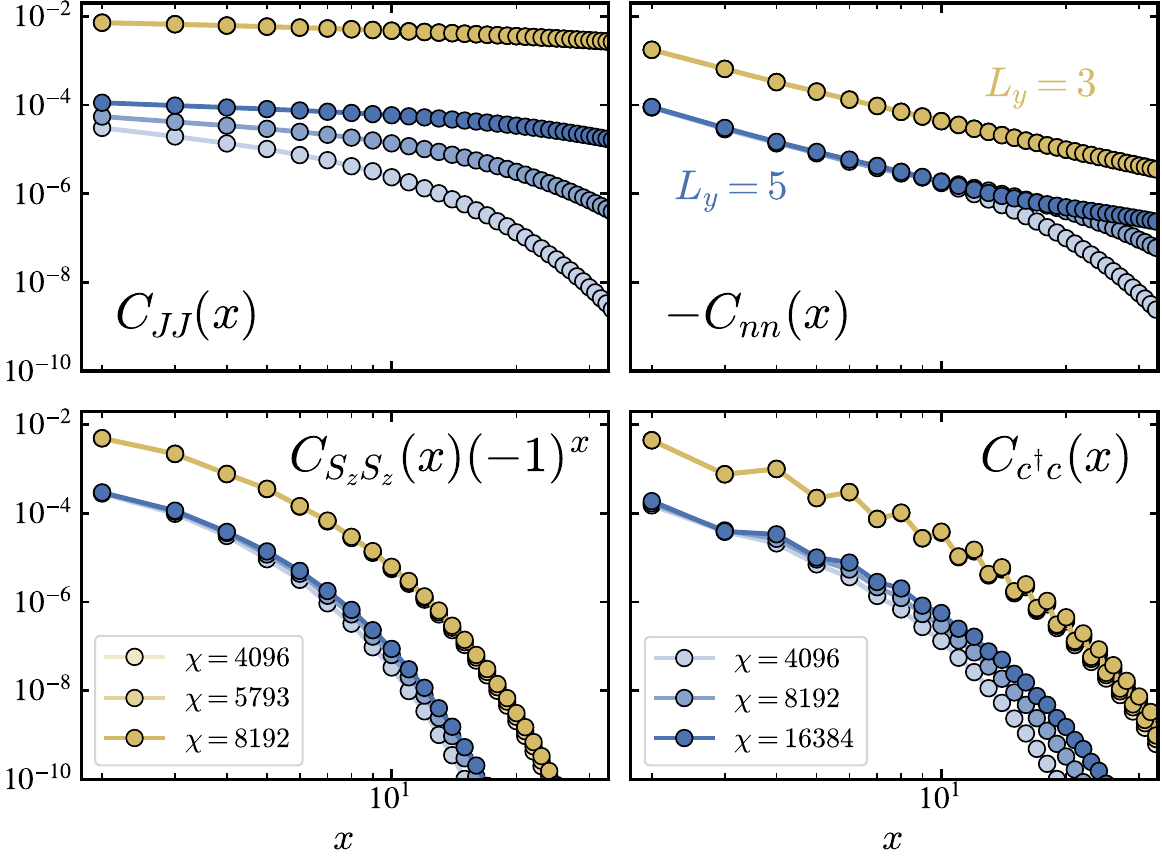}
\caption{Connected correlation functions on the odd $L_y$ cylinders at their critical point, $U_c(L_y=3)=11.1$ (yellow) and $U_c(L_y=5)=11.6$ (blue). The $L_y=5$ correlations are reduced by $\times 10$ for ease of viewing.
Upper: Current-current correlations and negative of density-density correlations, approaching quasi-long range order as a function of MPS bond dimension $\chi$ (increasing light to dark).
Lower: Spin-spin (plotted with alternating sign $(-1)^x$) and electron-hole correlations decay exponentially, demonstrating a spin gap.}
\label{fig:NeutralCorr}
\end{figure}

We now discuss the behavior of correlation functions. The connected current correlator $C_{JJ}(x) = \langle J_0^\perp J_x^\perp\rangle_\text{c}$ exhibits power-law decay at the approximate transition points $U_c(L_y=3)=11.1$ and $U_c(L_y=5)=11.6$ (Fig.~\ref{fig:NeutralCorr}a), chosen as $U = U_c$ where the correlation length is largest at the largest available $\chi$. The critical fluctuations also manifest in the charge density: the connected correlations $C_{nn}(x) = \langle n_0 n_x\rangle_\text{c}$ (with $n_x$ averaged over ring $x$) are negative and decay algebraically  (Fig.~\ref{fig:NeutralCorr}b), but fall off much more rapidly than the current correlations.
While there is no SSB for $L_y=4,6$ we still observe an enhanced correlation length (see Fig.~\ref{fig:OP_and_xi}) and slow decay of the current and density correlations~\cite{SoM}.

On the other hand, the connected spin correlations $C_{S^zS^z}(x) = \langle S^z_0 S^z_x \rangle_\text{c}$ (with $S^z_x$ averaged over ring $x$) exhibit clear exponential decay at the odd-$L_y$ critical points, oscillating in sign with a period of two rings, with no apparent enhancement from $L_y=3$ to $L_y=5$ (Fig.~\ref{fig:NeutralCorr}c). The electron correlators $C_{c^\dag c}(x) = L_y^{-2}\sum_{k_y} |\langle c^\dag(0,k_y) c(x,k_y) \rangle|$ likewise decay exponentially (Fig.~\ref{fig:NeutralCorr}d), so that both spin and electron excitations are gapped at the transition. Should these behaviors persist in the thermodynamic limit, they imply that the 2+1D critical point is continuous and hosts gapless, charge-neutral spin-singlet fluctuations, namely of the current.

We comment briefly on the nature of the odd-$L_y$ quasi-$(1+1)$D critical points. As the $1e$ and $2e$ correlators remain gapped (see Supplemental Material Sec.~G~\cite{SoM}), the data are incompatible with Luttinger and Luther-Emery liquid criticality~\cite{Giamarchibook,LutherEmery1974}. The spontaneous breaking of a $\mathbb{Z}_2$ symmetry suggests the transition is in the 2D Ising universality class. The ``ionic'' Hubbard chain~\cite{Nagaosa1986_1,Nagaosa1986_2,Egami1993} shares symmetries and behavior with the studied odd $L_y$ systems~\cite{FabrizioPRL,FabrizioNPB,Tincani2009,Verresen2021} and has a Mott transition; while the model has subtleties that have complicated the unambiguous determination of exponents, it is likely that the transition is of the Ising type (see Supplemental Material Sec.~G for a more detailed discussion~\cite{SoM}).
Further investigation into the odd $L_y$ sequence of transitions, particularly precisely how they evolve to the 2+1D transition and which critical correlations survive, is needed. The prominence of current-current correlations, namely their slow falloff at both short and long distances, suggests they remain prominent at the 2+1D critical point.

\smallskip
\textit{Discussion---}Our numerical data on both even and odd-circumference cylinder geometries provides confirming evidence of a CSL phase and indicates that the CSL-IQH transition is in fact continuous in the Hofstadter-Hubbard model at $\pi/2$ flux per triangle. The single-electron and spin gaps remain nonzero across the transition, while the current-current correlation function exhibits power-law behavior at the odd-$L_y$ critical points and is also prominent on the even-$L_y$ geometries. This raises the possibility of experimental detection of the transition and provides valuable guidance for its further numerical study. Using quantum Monte Carlo, fermionic PEPS, or neural quantum states, one may be able to access wider systems, yielding confirming evidence of quantum criticality, estimates of critical exponents, and possible signatures of conformal invariance~\cite{zhou2025_CSconfinement}. Our results support Ref.~\cite{Divic_2025PNAS}, which proposes a field theory of the 2+1D transition featuring critical fluctuations of the low-energy conserved current.

A broader message of this work is that the ability to extrapolate 2+1D physics from dimensionally-reduced data is greatly enhanced by analyzing how symmetries, ground states and their phase transitions respond to spatial compactification and flux threading.
In particular, the properties of transitions into topologically-ordered phases may generally be illuminated by studying cylinder compactifications in which topological ground state degeneracy manifests as SSB~\cite{TaoThouless1983,SeidelMoore2005}.
Leveraging such techniques systematically, it would be valuable to characterize the contexts in which UV features of Mott and quantum Hall phases offer improved understanding of their charge fluctuation-driven quantum phase transitions.

\begin{acknowledgments}
\textit{Acknowledgements---}We acknowledge useful discussions with E. Altman, S. Chatterjee, O. Gauth\'e, Y-C. He, M. Knap, C. Kuhlenkamp, J.Y. Lee, C. Liu, F. Pollmann, R. Sahay, R. Verresen, Z. Weinstein, Y-H. Zhang, and especially X-Y. Song and A. Vishwanath.
S.D. acknowledges support from the NSERC PGSD fellowship.
A.J.M. was supported in part by Programmable Quantum Materials, an Energy Frontier Research Center funded by the U.S. Department of Energy (DOE), Office of Science, Basic Energy Sciences (BES), under award DE-SC0019443.
M.Z. and S.D. were supported by the U.S. Department of Energy, Office of Science, National Quantum Information Science Research Centers, Quantum Systems Accelerator (QSA).
This research is funded in part by the Gordon and Betty Moore Foundation’s EPiQS Initiative, Grant GBMF8683 to T.S.
Calculations were performed using the TeNPy Library~\cite{tenpy,tenpy_v1,tenpy_v1_codebase}. 
The Flatiron Institute is a division of the Simons Foundation.
\end{acknowledgments}

\bibliographystyle{apsrev4-2} 

\bibliography{main} 



\definecolor{shadecolor}{gray}{0.9}

\clearpage

\onecolumngrid
\begin{center}
\textbf{\large Supplemental Material: ``Chiral Spin Liquid and Quantum Phase Transition in the Triangular Lattice Hofstadter-Hubbard Model''}
\end{center}

\setcounter{equation}{0}
\setcounter{figure}{0}
\setcounter{table}{0}
\makeatletter
\renewcommand{\thefigure}{S\arabic{figure}}
\renewcommand{\thesection}{S\Roman{section}}
\renewcommand{\thesubsection}{S\Roman{subsection}}
\renewcommand{\bibnumfmt}[1]{[S#1]}
\setcounter{section}{0}




\newcommand{\Danfigscale}{0.75}
\newcommand{\otherfigscale}{0.5}

\appendix

\tableofcontents

\section{Hamiltonian construction at all rational fluxes}

\subsection{YC Lattice Parameterization}\label{sec:YCgeo}

We first describe two coordinate parameterizations of the {underlying triangular lattice}, beginning in the 2D limit. In the ``YC'' parameterization of the 2D plane, the $\hat{y}$ site-to-site distance is the nearest-neighbour distance $a$. This corresponds to the Bravais vectors
\begin{align} \label{eq:underlyingYC_bravais_vecs}
    \bm{a}_{1}=a\left(\frac{\sqrt{3}}{2}\hat{x}+\frac{1}{2}\hat{y}\right), \qquad \bm{a}_{2}=a\hat{y} \qquad \text{(YC)},
\end{align}
and the corresponding reciprocal lattice vectors
\begin{align} \label{eq:underlyingYC_recip_vecs}
    \bm{b}_{1}=\frac{4\pi}{\sqrt{3}a}\hat{x}, \qquad \bm{b}_{2}=\frac{2\pi}{a}\left(-\frac{\hat{x}}{\sqrt{3}} + \hat{y}\right) \qquad \text{(YC)}.
\end{align}
This leads to a natural cylinder compactification: one may identify each point
\begin{align} \label{eq:site_coords}
    \bm{r} = x \bm{a}_1 + y\bm{a}_2
\end{align}
with points $\bm{r}'$ for which $y \equiv y'\ (\text{mod }L_y)$ and also $x=x'$. We refer to $L_y$ as the ``circumference'' of the cylinder. Doing so for the YC lattice, one obtains the canonical ``YC-$L_y$'' cylinder geometry~\cite{Szasz2020}, whose circumference is $L_y a$.

\subsection{Choosing the magnetic unit cell}

Let the flux per {parallelogram} (\textit{i.e.}, two triangles) be
\begin{align} \label{eq:flux}
    \Phi = \frac{p}{q} \Phi_{0},
\end{align}
where $\Phi_{0}=\frac{2\pi\hbar}{e} \sim 2\pi$ in units where $\hbar/e=1.$ The above definition of flux is convenient for defining the geometry, but we briefly remark that the flux $\Phi/2$ per triangle --- which is the smallest traversable plaquette --- is the quantity with respect to which physical behavior (\textit{e.g.}, the Hofstadter spectrum) must be $\Phi_0$-periodic. In any case, the corresponding magnetic field strength is
\begin{align} \label{eq:B_strength}
B = \frac{\Phi}{A_\text{par}} = \frac{p\Phi_{0}/q}{\frac{\sqrt{3}}{2}a^{2}}=\frac{2p\Phi_{0}}{\sqrt{3}q}a^{-2}.
\end{align}

Regardless of the parity of $q$, there will be $q$ magnetic sublattice sites in the magnetic unit cell.
When $q$ is \textbf{even}, the magnetic Bravais vectors are
\begin{align} \label{eq:magnetic_Bravais}
\bm{R}_{1} = q\bm{a}_{1}-\frac{q}{2}\bm{a}_{2} \propto \hat{x}, \qquad \bm{R}_{2} = \bm{a}_{2} \propto \hat{y}, \qquad (q\ \text{even}),
\end{align}
whose total area $A_{\text{muc}}$ is precisely that of $2q$ triangles of area $\frac{\sqrt{3}}{4}a^{2}.$
The associated reciprocal lattice vectors are
\begin{align} \label{eq:magnetic_reciprocal}
    \bm{G}_1 = \frac{1}{q}\bm{b}_1 \propto \hat{x}, \qquad \bm{G}_2 = \frac{1}{2}\bm{b}_1 + \bm{b}_2 \propto \hat{y}, \qquad (q\ \text{even}).
\end{align}
As a sanity check, the total of the magnetic Brillouin zone (mBZ) is then 
\begin{align}
A_{\text{mBZ}} = |\bm{G}_1\times\bm{G}_2| = (2\pi)^{2}\frac{2}{q\sqrt{3}a^{2}} = \frac{(2\pi)^{2}}{A_{\text{muc}}}.\label{eq:A_mBZ}
\end{align}

When $q$ is \textbf{odd}, the enclosing magnetic unit cell again contains exactly $2q$ triangles, but we take
\begin{align}
\bm{R}_{1} = q\bm{a}_{1}-\frac{q-1}{2}\bm{a}_{2}, \qquad \bm{R}_{2} = \bm{a}_{2} \propto \hat{y},  \qquad (q\ \text{odd}),
\end{align}
whose total area is once again
\begin{align}
    |\bm{R}_1\times \bm{R}_2| = q |\bm{a}_1\times \bm{a}_2| = \frac{q\sqrt{3}}{2}a^{2} = A_\text{muc}.
\end{align}
The reciprocal lattice vectors are now
\begin{align}
    \bm{G}_1 = \frac{1}{q}\bm{b}_1, \qquad \bm{G}_2 = \frac{q-1}{2q}\bm{b}_1 + \bm{b}_2, \qquad (q\ \text{odd}),
\end{align}
which also satisfy the inverse area relationship, Eq.~\eqref{eq:A_mBZ}. For both even and odd denominator $q,$ note that the enclosing magnetic unit cell contains exactly $2q$ triangles, so $2\pi p \in 2\pi \mathbb{Z}$ total flux.

\subsection{Hamiltonian and choice of vector potential}

In all cases, we adopt the standard convention for the on-site Hubbard repulsion:
\begin{align}
    H_\mathrm{int} = U \sum_{\bm{r}} n_{\bm{r}\uparrow}n_{\bm{r}\downarrow},\qquad U\geq0.
\end{align}
Now we discuss the hoppings. Placing the underlying triangular lattice on the cylinder as described in Sec.~\ref{sec:YCgeo}, let the resulting system have circumference $L_{y}$, \textit{i.e.}, the points $\bmr \sim \bmr + L_y\bma_2$ are geometrically identified. Let the lattice sites be indexed as in Eq.~\eqref{eq:site_coords}, which defines a set of coordinate coefficients $(x,y)\in\mathbb{Z}^{2}.$ Since $\bma_{2}$ is purely vertical, then $x$ indexes the cylinder ``rings'', and $y$ is an integer-valued spatial index within each ring.
To generate complex hoppings and modify the translation algebra into a magnetic translation algebra, we introduce an external $U(1)$ vector potential. We assume that it is linear and that its curl is the magnetic field:
\begin{align} \label{eq:generic_vectorpotential}
    \bm A(\bm{r}) = M \bm{r},\qquad \nabla\times \bm A(\bm{r}) = B\hat{z},
\end{align}
where $B$ is as in Eq.~\eqref{eq:B_strength}. The hoppings are obtained from the vector potential via the Peierl's substitution, for which we adopt the following sign convention:
\begin{align} \label{eq:linearA_hoppings}
    t_{\bmr'\leftarrow \bmr} = t\exp\left(i\int^{\bmr'}_{\bmr} \bm{A}(\bm{\ell})\cdot d\bm{\ell} \right) = t\exp\left(i \bm{A}\left(\frac{\bmr'+\bmr}{2}\right)\cdot (\bmr'-\bmr)\right),
\end{align}
where the second equality is due to linearity of $\bm{A}(\bmr)$. Our convention is that the single-particle Hamiltonian is (note the minus sign):
\begin{align} \label{eq:hopping_ham_arrows}
    \hat{h} = -\sum_{\bm{r}',\bm{r}}c^{\dagger}(\bm{r}')t_{\bmr'\leftarrow\bmr}c(\bm{r}),
\end{align}
where we only include nearest-neighbour hopping on the triangular lattice, for which
\begin{align} \label{eq:Peierls_hoppings}
    t_{\bmr'\leftarrow\bmr} = te^{i\bm{A}\left(\frac{\bm{r}'+\bm{r}}{2}\right)\cdot(\bm{r}'-\bm{r})}.
\end{align}

To constrain the form of the vector potential, we remark that the terms $x\hat{x}$ and $y\hat{y}$ are curl-free, whereas $Bx\hat{y}$ and $-By\hat{x}$ (the standard Landau gauge potentials on the rectangular lattice) have the correct curl. We therefore reason that the most general permissible $M$ in Eq.~\eqref{eq:generic_vectorpotential} is given by
\begin{align}
    M = B \begin{pmatrix} \alpha & \gamma-1 \\ \gamma & \beta \end{pmatrix},
\end{align}
where $\alpha,\beta,\gamma \in \R$. We will now demand that the choice of vector potential is compatible with our chosen geometry of magnetic Bravais vectors. Specifically, we ask that the Peierl's hoppings generated by $\bm{A}$ are \textit{periodic} in the chosen magnetic Bravais vectors. This allows us to place the system easily on a torus or cylinder, provided the system is an integer tiling of magnetic unit cells.
Quantitatively, we demand that the RHS of Eq.~\eqref{eq:linearA_hoppings} is invariant under shifts of the argument of $\bm{A}$ by the magnetic Bravais vectors $\bm{R}_1$ and $\bm{R}_2$, for all bond vectors $\bmr'-\bmr$. By linearity of the vector potential, and since the bond vectors are linear combinations of $\bm{a}_1$ and $\bm{a}_2$, it suffices to require that
\begin{align}
    \bm{a}_i \cdot \bm{A}(\bm{R}_j) \in 2\pi\Z,
\end{align}
for all $i,j$. Following Ref.~\cite{HerzogArbeitman2020}, an elegant solution can be found in terms of the reciprocal vectors of the underlying triangular lattice. We take the following since $\bm{b}_{1}\propto\hat{x}$, and we want the vector potential to be independent of $y$:
\begin{align}
    \bm{A}(\bm{r})=\frac{\Phi}{(2\pi)^{2}}(\bm{r}\cdot\bm{b}_{1}) \bm{b}_{2},
\end{align}
where $\Phi$ is the flux per \textit{parallelogram}.
Noting the vector identity
\begin{align}
\left[\nabla\times\left((\bm{r}\cdot\bm{Q})\bm{P}\right)\right]_{i}&=\epsilon_{ijk}\partial_{j}\left(\bm{r}_{\ell}\bm{Q}_{\ell}\bm{P}_{k}\right) =\epsilon_{ijk}\bm{Q}_{j}\bm{P}_{k} =(\bm{Q}\times\bm{P})_{i},
\end{align}
then
\begin{align}
\bm{B}(\bm{r})&=\nabla\times\bm{A}(\bm{r}) = \frac{\Phi}{(2\pi)^{2}}\bm{b}_{1} \times \bm{b}_{2} = \frac{\Phi}{(2\pi)^{2}}\frac{(2\pi)^{2}}{A_{\text{par}}}\hat{z} = \frac{\Phi}{A_{\text{par}}}\hat{z}.
\end{align}
Moreover, it can be shown to have the required periodicity property. Note that
\begin{align}
\bm{a}_{i}\cdot\bm{A}(\bm{R}_{j})&=\frac{\Phi}{(2\pi)^{2}}(\bm{R}_{j}\cdot\bm{b}_{1})\bm{b}_{2} \cdot\bm{a}_{i} = \delta_{i,2} \frac{2\pi p/q}{2\pi}(\bm{R}_{j}\cdot\bm{b}_{1}) = \delta_{i,2} \frac{p}{q}\bm{R}_{j}\cdot\bm{b}_{1},
\end{align}
which clearly vanishes when $j=2.$ Otherwise,
\begin{align}
\bm{a}_{i}\cdot\bm{A}(\bm{R}_{1})&= \delta_{i,2} \frac{p}{q}\left(\begin{cases}
\left(q\bm{a}_{1}-\frac{q}{2}\bm{a}_{2}\right)\cdot\bm{b}_{1} & q\text{ even}\\
\left(q\bm{a}_{1}-\frac{q-1}{2}\bm{a}_{2}\right)\cdot\bm{b}_{1} & q\text{ odd}
\end{cases}\right)= 2\pi p \delta_{i,2},
\end{align}
so that, indeed, in either case:
\begin{align}
\bm{a}_{i}\cdot\bm{A}(\bm{R}_{j})\in2\pi\mathbb{Z}.
\end{align}

\begin{figure}[h]
    \centering
    \includegraphics[width = 494 pt]{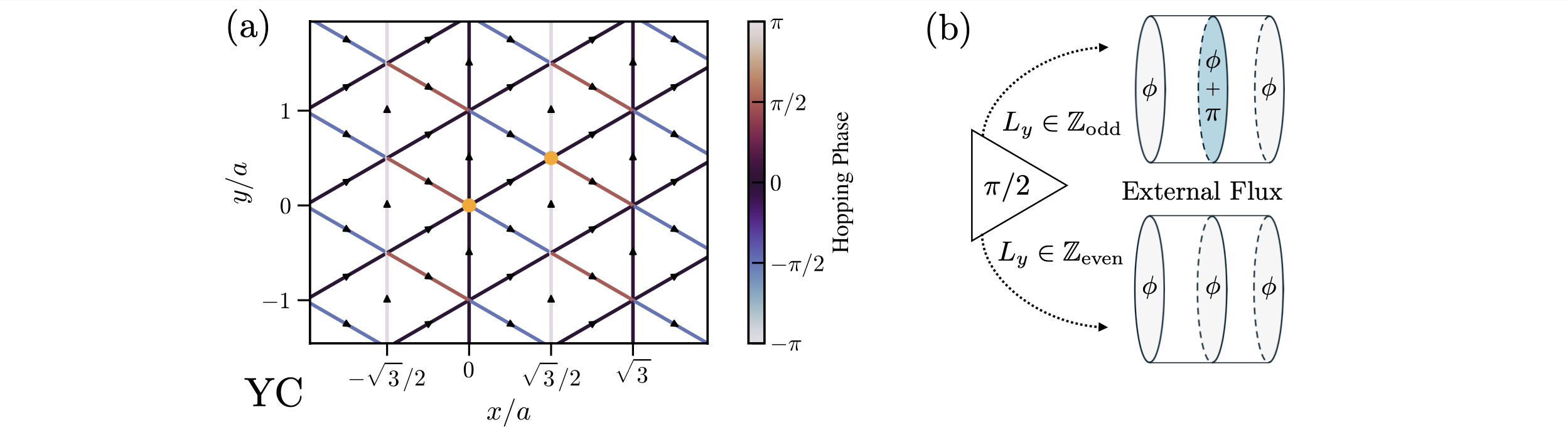}
    \caption{(a) Graphical depiction of the complex phases of the nearest-neighbor hopping amplitudes of Eq.~\eqref{eq:Peierls_hoppings} for the $(p,q)=(1,2)$ two-band Hamiltonian in a portion of the infinite \textit{2D plane}, where the orientation of the plane is consistent with the YC coordinate system specified by Eqs.~(\ref{eq:underlyingYC_bravais_vecs},\ref{eq:magnetic_Bravais}).
    The color of each bond denotes the complex phase of $t_{\bm{r}'\leftarrow \bm{r}}$ (see Eq.~\eqref{eq:hopping_ham_arrows}) in the direction of the arrow (pictured at the bond centers).
    The two orange dots indicate the two sites in the magnetic unit cell. Note that the top/bottom and right/left edges are \textit{not} implicitly identified.
    (c) Graphical depiction of the fluxes through adjacent loops of a YC-$L_y$ cylinder, defined in Sec.~\ref{sec:YCgeo}. The outside surface of each cylindrical segment has a total area of $2L_y$ triangles, each having $\pi/2$ flux [pictured]. When $L_y$ is odd [upper scenario], this leads to a $\pi$-staggering of the flux through each cylinder ring due to Gauss' law. When $L_y$ is even [lower scenario], the flux through each ring is the same.}
    \label{fig:SM_lattice}
\end{figure}

Specializing to the case $(p,q)=(1,2)$ relevant to the main text, we plot the resulting hopping amplitudes Eq.~\eqref{eq:Peierls_hoppings} in Fig.~\ref{fig:SM_lattice}(a) for the YC parameterization of the 2D lattice. The hoppings only take values in $\{\pm 1, \pm i\}$ and are evidently periodic in both $\bm{R}_1$ and $\bm{R}_2$.
The color of each bond denotes the complex phase of $t_{\bm{r}'\leftarrow \bm{r}}$ (see Eq.~\eqref{eq:hopping_ham_arrows}) in the direction of the arrow (pictured at the bond centers).

\section{Threading flux}

\subsection{Affine-linear vector potential implementation}

When dimensionally-reducing the lattice system to the cylinder, the charge and spin flux through the cross sections of the cylinder distinguish gauge-inequivalent systems. This additional flux can be implemented by adding a constant offset to the vector potential:
\begin{align} \label{eq:vectorpot_fluxthread}
    \bm{A}(\bm{r}) \longrightarrow \bm{A}_{\phi}(\bm{r}) = \bm{A}(\bm{r}) + \phi \frac{\bm{b}_2}{2\pi L_y},
\end{align}
where
\begin{align}
    \phi = \phi_q + \frac{\sigma_z}{2}\phi_s.
\end{align}
This modifies the hopping elements by
\begin{align} \label{eq:hopping_after_threading}
    t_{\bm{r}+\bm{\delta}\leftarrow\bm{r}}\longrightarrow t_{\bm{r}+\bm{\delta}\leftarrow\bm{r}}^{\phi}=t_{\bm{r}+\bm{\delta}\leftarrow\bm{r}} \exp\left(i \frac{\phi}{2\pi} \frac{\bm{b}_2 \cdot \bm{\delta}}{L_y}\right).
\end{align}
Since we made the choice $\bm{A} \propto \bm{b}_2$ for the original vector potential (in the absence of flux threading), then this is equivalent to a shift in the origin:
\begin{align}
    \bm{A}_{\phi}(\bm{r})=\bm{A}(\bm{r}+\bm{X}_\phi)=\bm{A}(\bm{r}) + \frac{\Phi}{(2\pi)^{2}}(\bm{X}_\phi\cdot\bm{b}_{1})\bm{b}_{2},
\end{align}
where
\begin{align} \label{eq:translation_for_flux}
    \bm{X}_\phi = \frac{\phi \bm{a}_1}{\Phi L_{y}}.
\end{align}

\subsection{Twisted boundary condition implementation}

For the purposes of numerical implementation, particularly in hybrid space where we trade $y$ for $k_y$, it is convenient to instead implement flux threading via a twist in the boundary conditions. Starting with periodic boundary conditions for the fermion operators at $\phi=0$, namely $c^{\dagger}(\bm{r} + L_y\bm{a}_2) = c^{\dagger}(\bm{r})$, we define a new set of operators
\begin{align} \label{eq:cdag_redef_flux}
    c^\dag_\phi(\bm{r}) = c^\dag(\bm{r}) \exp(i y\phi/L_y) = c^\dag(\bm{r}) \exp\left(i \phi \frac{\bm{b}_2 \cdot \bm{r}}{2\pi L_y} \right).
\end{align}
Then instead of implementing Eq.~\eqref{eq:vectorpot_fluxthread}, we can instead replace every instance of $c$ in the second-quantized interacting Hamiltonian with $c_\phi$. This does not affect terms that depend solely on charge and spin densities, \textit{e.g.}, the Hubbard interaction. However, it modifies the hoppings in a way that is equivalent to Eq.~\eqref{eq:vectorpot_fluxthread}. Under the operator replacement, the hoppings are replaced by
\begin{align}
    c^\dag(\bmr + \bm{\delta}) t_{\bmr+\bm{\delta}\leftarrow \bmr} c(\bmr) &\longrightarrow c_\phi^\dag(\bmr + \bm{\delta}) t_{\bmr+\bm{\delta}\leftarrow \bmr} c_\phi(\bmr) \\
    &= c^\dag(\bmr + \bm{\delta}) \exp\left(i \phi \frac{\bm{b}_2 \cdot (\bm{r} + \bm{\delta})}{2\pi L_y} \right) t_{\bmr+\bm{\delta}\leftarrow \bmr} \exp\left(-i \phi \frac{\bm{b}_2 \cdot \bm{r}}{2\pi L_y} \right) c(\bmr) \\
    &= c^\dag(\bmr + \bm{\delta}) \exp\left(i \frac{\phi}{2\pi} \frac{\bm{b}_2 \cdot \bm{\delta}}{L_y} \right) t_{\bmr+\bm{\delta}\leftarrow \bmr} c(\bmr),
\end{align}
which is exactly equivalent to Eq.~\eqref{eq:hopping_after_threading}. To see that Eq.~\eqref{eq:cdag_redef_flux} corresponds to a twisting of the boundary condition on the fermion operators, we note that
\begin{align} \label{eq:operator_twistbc}
    c^\dag_\phi(\bm{r} + L_y\bma_2) = c^\dag(\bm{r} + L_y\bma_2) \exp\left(i \phi \frac{\bmb_2\cdot(\bmr + L_y\bma_2)}{2\pi L_y} \right) = c_\phi^\dag(\bm{r}) \exp(i \phi).
\end{align}

The utility of this alternative formulation is that it leads to a clean numerical implementation upon fourier transforming $y \to k_y$. In particular, note that $(x,y) = (\bmr\cdot\bmb_1/2\pi, \bmr\cdot\bmb_2/2\pi)$ and define
\begin{align} \label{eq:hybridFourier}
    c_\phi^{\dagger}(x,k_{y}) = \frac{1}{\sqrt{L_{y}}}\sum_{y}e^{-ik_{y}y}c_\phi^{\dagger}(x,y)\iff c_\phi^{\dagger}(x,y) = \frac{1}{\sqrt{L_{y}}}\sum_{k_{y}}e^{ik_{y}y}c_\phi^{\dagger}(x,k_{y}).
\end{align}
Note that this is the Fourier convention of TeNPy. When we express the Hamiltonian in the hybrid $xk_{y}$ representation, we find that all that is required is to shift the allowed values of $k_{y}.$ To see this, we impose the twisted boundary condition Eq.~\eqref{eq:operator_twistbc} on the Fourier-expanded operators:
\begin{align}
    \frac{1}{\sqrt{L_{y}}}\sum_{k_y}e^{ik_y(y+L_{y})}c_{\theta}^{\dagger}(x,k_y)=\frac{1}{\sqrt{L_{y}}}\sum_{k_y}e^{ik_y y}c_{\theta}^{\dagger}(x,k_y)e^{i\phi},
\end{align}
so that
\begin{align}
    e^{ik_y L_y}=e^{i\phi}.
\end{align}
The allowed values of momenta are therefore quantized to
\begin{align} \label{eq:shifted_k}
    k_y = \frac{2\pi}{L_{y}}\left(n + \frac{\phi_q}{2\pi} + \frac{\sigma_{z} \phi_s}{4\pi}\right), \qquad n\in\mathbb{Z},
\end{align}
which are shifted relative to the $\phi=0$ case. When the original Hamiltonian $\hat{H}_{\phi=0}$ is written in hybrid space, some coefficients will depend on $k.$ The only effect of the replacement Eq.~\eqref{eq:cdag_redef_flux} is then for those coefficients to instead be evaluated on this shifted set of allowed values~\cite{Szasz2020}.

\section{Two-band hopping Hamiltonian}
\label{sec:two-band-hopping-Hamiltonian}

\begin{figure}[h]
    \centering
    \includegraphics[width = 494 pt]{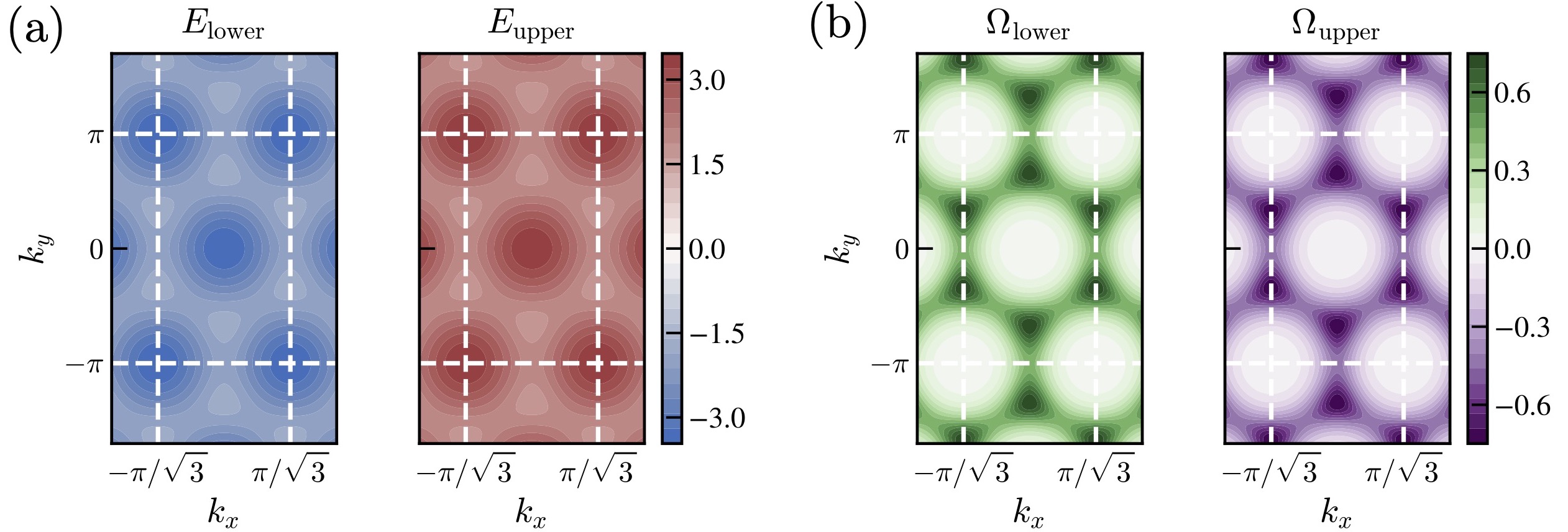}
    \caption{(a) Plot of the lower and upper energy bands of the Hofstadter system with $(p,q)=(1,2)$, corresponding to the case with $\Phi_\triangle=\pi/2$ considered throughout the main text. Momenta are given within the first ``magnetic'' Brillouin zone, defined with respect to the magnetic reciprocal vectors of Eq.~\eqref{eq:magnetic_reciprocal} in the 2D planar YC coordinate system of Eq.~\eqref{eq:underlyingYC_recip_vecs}.
    (b) Plot of the Berry curvature of the aforementioned bands which, like their energy, is equal and opposite at every momentum. The bands carry opposite Chern numbers $\pm 1$.}
    \label{fig:SM_bands_berry}
\end{figure}

\subsection{Hamiltonian at $(p,q)=(1,2)$}

Now let us specialize to the YC lattice geometry and to the case $(p,q)=(1,2)$, which corresponds to $\Phi_\triangle = \pi/2$. To specify the unit cells (which each contain two sites), we write $x=\tau+2u$, where $u\in\mathbb{Z}$, while $\tau\in\{0,1\}$ specifies the even/odd rings. With periodic boundary conditions and absent any flux insertion, the hoppings Eq.~\eqref{eq:Peierls_hoppings} take the particular form
\begin{align}\label{eq:hopping_YC_up}
t_{(x,y+1)\leftarrow(x,y)}&=t\exp\left(i\bm{A}\left(x\bm{a}_{1}+y\bm{a}_{2}+\frac{\bm{a}_{2}}{2}\right)\cdot\bm{a}_{2}\right) \nonumber\\
&=t\exp\left(\frac{i}{4\pi}\left(x\bm{a}_{1}+y\bm{a}_{2}+\frac{\bm{a}_{2}}{2}\right)\cdot\bm{b}_{1}\bm{b}_{2}\cdot\bm{a}_{2}\right) \nonumber\\
&=t\exp\left(i\pi x\right),
\end{align}
\begin{align}\label{eq:hopping_YC_rightup}
t_{(x+1,y)\leftarrow(x,y)}&=t\exp\left(i\bm{A}\left(\frac{\bm{a}_{1}}{2}+x\bm{a}_{1}+y\bm{a}_{2}\right)\cdot\bm{a}_{1}\right) \nonumber \\
&=t,
\end{align}
\begin{align}\label{eq:hopping_YC_rightdn}
t_{(x+1,y-1)\leftarrow(x,y)}&=t\exp\left(i\bm{A}\left(\frac{\bm{a}_{1}-\bm{a}_{2}}{2}+x\bm{a}_{1}+y\bm{a}_{2}\right)\cdot(\bm{a}_{1}-\bm{a}_{2})\right) \nonumber\\
&=t\exp\left(-i\bm{A}\left(\frac{\bm{a}_{1}-\bm{a}_{2}}{2}+x\bm{a}_{1}+y\bm{a}_{2}\right)\cdot\bm{a}_{2}\right) \nonumber\\
&=t\exp\left(-i\pi(x+1/2)\right),
\end{align}
as shown graphically in Fig.~\ref{fig:SM_lattice}(a).
We depict the resulting energy bands $E_\text{lower/upper}(\bm{k})$ in Fig.~\ref{fig:SM_bands_berry}, as well as their corresponding Berry curvature $\Omega_\text{lower/upper}(\bm{k})$, in the YC coordinate system consistent with Eqs.~\eqref{eq:underlyingYC_recip_vecs} and \eqref{eq:magnetic_reciprocal}. The energies are equal and opposite at every momentum and are gapped everywhere in the Brillouin zone. The Berry curvature is also everywhere equal and opposite, with the Chern numbers evaluating to $+1$ for the lower band and $-1$ for the upper band.

\subsection{Symmetries in the 2D plane}

\subsubsection{Magnetic translations}

Recall that we designed the vector potential so that the Hamiltonian would be invariant under $T_{y}$ translations, \textit{i.e.}, in the circumferential direction $\bm{a}_{2},$ defined as
\begin{align}
T_{y}c_{\bm{r}}^{\dagger}T_{y}^{-1}=c_{\bm{r}+\bm{a}_{2}}^{\dagger}.
\end{align}
To see this, note that
\begin{align}
t_{\bm{r}+\bm{a}_{2}+\bm{\delta}\leftarrow\bm{r}+\bm{a}_{2}}&=t\exp\left(i\bm{A} \left(\bm{r}+\bm{a}_{2}+\bm{\delta}/2\right)\cdot\bm{\delta}\right)\\&=t\exp\left(i \bm{A} \left(\bm{r}+\bm{\delta}/2\right) \cdot\bm{\delta}\right)\\&= t_{\bm{r}+\bm{\delta}\leftarrow\bm{r}},
\end{align}
since $\bm{A}(\bma_2)=0$ by design. On the other hand, translation along $\bm{a}_{1}$ introduces an additional term:
\begin{align}
t_{\bm{r}+\bm{a}_{1}+\bm{\delta}\leftarrow\bm{r}+\bm{a}_{1}} &= t\exp\left(i \bm{A}\left(\bm{a}_{1} + \bm{r}+\bm{\delta}/2\right)\cdot\bm{\delta}\right)\\&=\exp\left(i\bm{A}\left(\bm{a}_{1}\right)\cdot\bm{\delta}\right)t_{\bm{r}+\bm{\delta}\leftarrow\bm{r}} \\
&=\exp\left(\frac{i}{2}\bm{b}_{2}\cdot\bm{\delta}\right)t_{\bm{r}+\bm{\delta}\leftarrow\bm{r}}. \label{eq:a1_shift_t}
\end{align}
This affects the hoppings in directions with some weight along $\bm{\delta}=\bm{a}_{2}$ and, by comparison to Eq.~\eqref{eq:translation_for_flux} above, corresponds to threading $\phi=\pi L_{y}$ through the cylinder.
In the 2D plane geometry---and on cylinder geometries of even circumference as we'll later show---the Hamiltonian can be restored by composing the bare $T_{x}$ with a large gauge transformation. In particular, define the magnetic translation operation by
\begin{align} \label{eq:Tmag}
    T_{x} c_{\bm{r}}^{\dagger}T_{x}^{-1} = c_{\bm{r}+\bm{a}_{1}}^{\dagger}\exp\left(\frac{i}{2}\bm{b}_{2}\cdot\bm{r}\right) = c_{\bm{r}+\bm{a}_{1}}^{\dagger}(-1)^y,
\end{align}
so that
\begin{align}
    T_{x}\hat{h}T_{x}^{-1} &= T_{x}\left(\sum c_{\bm{r}+\bm{\delta}}^{\dagger}t_{\bm{r}+\bm{\delta}\leftarrow\bm{r}} c_{\bm{r}}\right) T_{x}^{-1} \\
    &= \sum\exp\left(\frac{i}{2}\bm{b}_{2}\cdot(\bm{r}+\bm{\delta})\right)c_{\bm{r}+\bm{a}_{1}+\bm{\delta}}^{\dagger}t_{\bm{r}+\bm{\delta}\leftarrow\bm{r}} c_{\bm{r}+\bm{a}_{1}}\exp\left(-\frac{i}{2}\bm{b}_{2}\cdot\bm{r}\right) \\
    &= \sum c_{\bm{r}+\bm{a}_{1}+\bm{\delta}}^{\dagger}\left(\exp\left(\frac{i}{2}\bm{b}_{2}\cdot\bm{\delta}\right)t_{\bm{r}+\bm{\delta}\leftarrow\bm{r}} \right)c_{\bm{r}+\bm{a}_{1}} \\
    &= \sum c_{\bm{r}+\bm{a}_{1}+\bm{\delta}}^{\dagger} t_{\bm{r}+\bm{a}_{1}+\bm{\delta}\leftarrow\bm{r}+\bm{a}_{1}} c_{\bm{r}+\bm{a}_{1}} \\
    &= \hat{h},
\end{align}
where the second-last equality follows from Eq.~\eqref{eq:a1_shift_t} above.

\subsubsection{Particle-hole}

Consider the most naive definition of the particle-hole operation:
\begin{align} \label{eq:naive_PH}
    \tilde{\mcP} c^\dag(\bmr) \tilde{\mcP}^{-1} = c(\bmr), \qquad \tilde{\mcP} i \tilde{\mcP}^{-1} = +i,
\end{align}
where the second condition specifies that it is a linear (as opposed to anti-linear) unitary many-body operator. If 1 and 2 denote an adjacent pair of sites, then this naive particle-hole operation acts as
\begin{align}
\tilde{\mathcal{\mathcal{P}}}(t_{12}c_{1}^{\dagger}c_{2}+t_{12}^{*}c_{2}^{\dagger}c_{1})\tilde{\mathcal{\mathcal{P}}}^{-1} & =t_{12}c_{1}c_{2}^{\dagger}+t_{12}^{*}c_{2}c_{1}^{\dagger}\\
 & =-t_{12}c_{2}^{\dagger}c_{1}-t_{12}^{*}c_{1}^{\dagger}c_{2}\\
 & =-t_{12}^{*}c_{1}^{\dagger}c_{2}-t_{12}c_{2}^{\dagger}c_{1}, \label{eq:PH_transform_hopping}
\end{align}
so that it reverses the Peierl's phase and shifts it by $\pi.$ On the triangular lattice, this modifies the flux through each elementary triangular plaquette as follows:
\begin{align} \label{eq:triangle_PH_action}
\Phi_{\triangle}\to-\Phi_{\triangle}+\pi.
\end{align}
Requiring that this operation leave the flux invariant modulo $2\pi,$ we find that we need
\begin{align}
2\Phi_{\triangle}=\pi\left(2n+1\right),\qquad n\in\mathbb{Z}.
\end{align}
This is consistent with the observation that the Hofstadter spectrum on the triangular lattice only has bands with opposite energy for these values of $\Phi_\triangle$. In particular, in this work we consider $\Phi_\triangle = \pi/2$. We find at each momentum $\bm{k}$ that both the energy and Berry curvature are exactly opposite, with the bands having Chern number $\pm 1.$

For a generic choice of Peierls' hopping phases consistent with this uniform magnetic flux, the naive particle-hole operation must be composed with a gauge transformation in order to exactly restore all the hopping elements. Given the particular form of the hoppings Eq.~\eqref{eq:hopping_YC_up} to Eq.~\eqref{eq:hopping_YC_rightdn} for the two-band model, one may easily confirm that the following form of the particle-hole operation leaves the Hamiltonian invariant:
\begin{align} \label{eq:spinless_PH_symmetry}
    \mcP c^\dag(\bmr) \mcP^{-1} = (-1)^{x+y} c^\mathsf{T}(\bmr), \qquad \mcP i \mcP^{-1} = +i,
\end{align}
where $\mathsf{T}$ is the transposition operation on the row vector of electronic annihilation operators (indexed by spin).

To obtain the form of the particle-hole symmetry presented in the main text, we compose the version in Eq.~\eqref{eq:spinless_PH_symmetry} with a conventional spin rotation:
\begin{align}
    \mcP c^\dagger_{x,y} \mcP^{-1} = (-1)^{x+y} c^\mathsf{T}_{x,y} (i\sigma_y), \quad \mcP i \mcP^{-1} = +i.
\end{align}
The spin rotation is chosen so that the spin operator
\begin{align}
    S_\alpha(\bmr) = \frac{1}{2}c^\dag(\bmr) \sigma_\alpha c(\bmr)
\end{align}
is left invariant. To see this, we note that the unitarity of $\mcP$ implies that $\mathcal{P}c(\bm{r})\mathcal{P}^{-1}=(-1)^{x+y}(-i\sigma_{y})c^{\dagger\mathsf{T}}(\bm{r})$. This allows us to explicitly compute:
\begin{align}
    \mathcal{P}S_{\alpha}(\bm{r})\mathcal{P}^{-1}&=\frac{1}{2}\mathcal{P}c^{\dagger}(\bm{r})\mathcal{P}^{-1}\sigma_{\alpha}\mathcal{P}c(\bm{r})\mathcal{P}^{-1}=\frac{1}{2}c^{\mathsf{T}}(\bm{r})\sigma_{y}\sigma_{\alpha}\sigma_{y}c^{\dagger\mathsf{T}}(\bm{r}).
\end{align}
When $\alpha\in\{x,z\}$ we acquire a minus sign but then obtain another from fermion anti-commutation when undoing the transposes. On the other hand, when $\alpha=y$ the Pauli matrices commute but the minus sign from fermion anti-commutation cancels with the fact that $\sigma_y$ is an anti-symmetric matrix:
\begin{align}
    \mathcal{P}S_{y}(\bm{r})\mathcal{P}^{-1}=\frac{1}{2}c_{s}(\bm{r})(\sigma_{y})_{ss'}c_{s'}^{\dagger}(\bm{r})=-\frac{1}{2}c_{s'}^{\dagger}(\bm{r})(\sigma_{y})_{ss'}c_{s}(\bm{r}) = +S_{y}(\bm{r}).
\end{align}

\subsection{Fate of Symmetries on Cylinder}

Let us elucidate when placing the lattice on a YC-$L_y$ cylinder explicitly breaks particle-hole symmetry. We may use the naive definition of particle-hole Eq.~\eqref{eq:naive_PH} because the fluxes in question are gauge-invariant quantities. Focusing on a particular cylinder ring $r$ penetrated with $\Phi_{r}$ flux, note that
\begin{align}
\tilde{\mathcal{\mathcal{P}}}:\Phi_{r}\mapsto\begin{cases}
-\Phi_{r} & L_{y}\text{ even},\\
\pi -\Phi_{r} & L_{y}\text{ odd}.
\end{cases}
\end{align}
This follows from the application of Eq.~\eqref{eq:PH_transform_hopping} to each of the $L_y$ nearest-neighbor links.
Therefore, $\tilde{\mathcal{\mathcal{P}}}$ is a symmetry (up to a gauge transformation) when, for all rings $r$, we have that
\begin{align} \label{eq:PH_ringflux}
\Phi_{r}\in\begin{cases}
\pi\mathbb{Z} & L_{y}\text{ even},\\
\frac{\pi}{2}+\pi\mathbb{Z} & L_{y}\text{ odd}.
\end{cases}
\end{align}

When $L_{y}$ is even, Gauss's law dictates that the fluxes $\Phi_{r_1}$ and $\Phi_{r_2}$ through two adjacent rings $r_{1}$ and $r_{2}$ must be equal. Note that this allows $T_{x}$ to be a symmetry. Therefore, possibly after threading $\phi$ external flux through
each ring of the cylinder, it is possible to tune all the $\Phi_{r}$ simultaneously (to $0$ or $\pi$) so that $\tilde{\mathcal{\mathcal{P}}}$ is respected. Only in those cases is $\tilde{\mathcal{\mathcal{P}}} T_{x}$ also a symmetry (up to a gauge transformation).

On the other hand, when $L_{y}$ is odd, Gauss's law instead dictates that
\begin{align} \label{eq:oddLy_gauss}
\Phi_{r_{2}}-\Phi_{r_{1}}=\pi\qquad\text{mod }2\pi.
\end{align}
This always rules out $T_{x}$ as a symmetry. However, so long as
$\Phi_{r_{1}}$ is $\pi/2$ or $-\pi/2,$ both Eqs.~\eqref{eq:PH_ringflux} and \eqref{eq:oddLy_gauss} can be respected, so that $\tilde{\mathcal{\mathcal{P}}}$ is a symmetry (up to a gauge transformation). On the other hand, if we instead tune $\Phi_{r_{1}}$ to 0 or $\pi,$ then $\tilde{\mathcal{\mathcal{P}}}T_{x}$ is a symmetry (up to a gauge transformation) even though neither $T_{x}$ nor $\tilde{\mathcal{\mathcal{P}}}$ is individually a symmetry; for the $L_y=3,5$ cylinders in the main text, we choose $\Phi_r = 0$ for all even rings $r$ and $\Phi_r = \pi$ for all odd rings $r$, which makes these cylinders invariant under $\tilde{\mathcal{\mathcal{P}}}T_{x}$.

\section{Entanglement spectrum} \label{sec:SM_entanglement}

\begin{figure}[h]
    \centering
    \includegraphics[width = \textwidth]{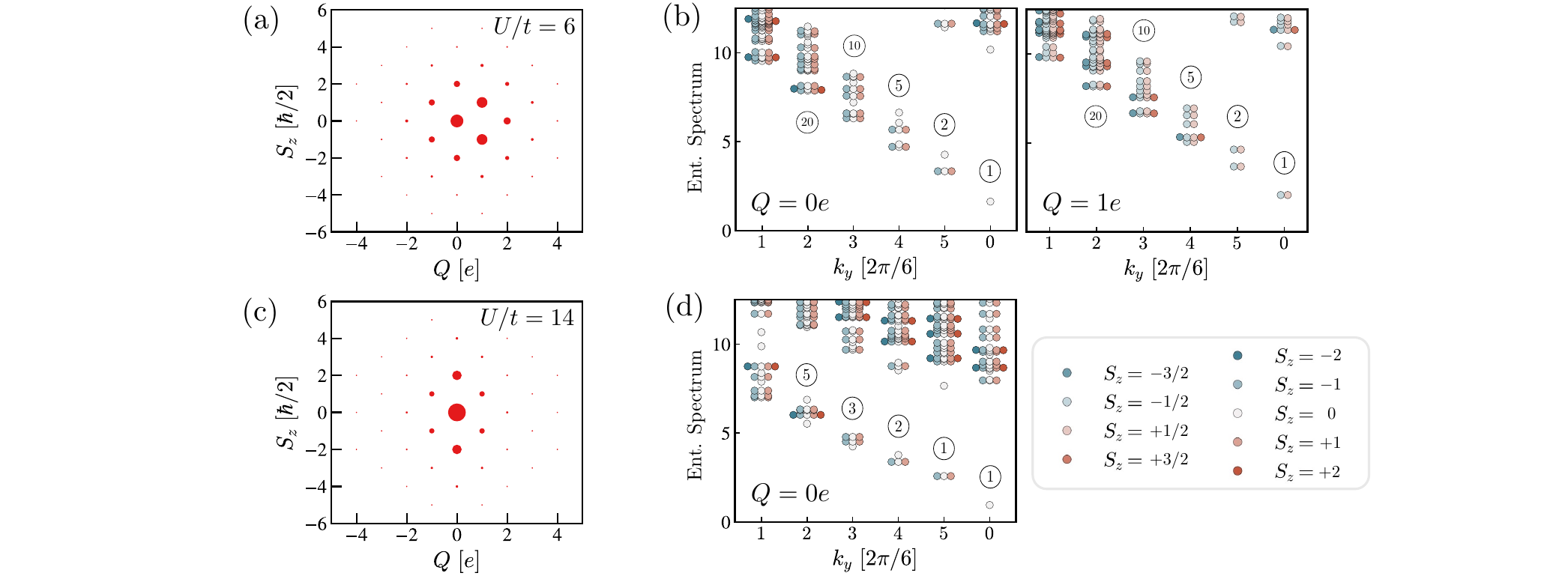}
    \caption{Entanglement data at $L_y=6$ with $\Phi_\text{ext}=\pi/2$ and $\chi=4096$, taken at the cut to the right of the $x=0$ ring. (a) Weight of the Schmidt values resolved by spin $S_z$ and charge $Q$ quantum numbers (see Appendix Sec.~\ref{sec:SM_entanglement} for details) at $U/t=6$. (b) Entanglement eigenvalues vs. circumferential momentum $k_y$; colors (with small artificial horizontal separation) indicate spin quantum number (see legend). Left panel: for $Q=0e$, with circled numerals indicating the number of nearly-degenerate $S_z=0$ entanglement eigenvalues at each momentum. Right panel: likewise for $Q=1e$, with numerals counting the number of nearly-degenerate $S_z=1/2$ states.
    (c,d) Same for $U/t=14$.}
    \label{fig:YC6_ES}
\end{figure}

Here we study the real-space entanglement spectrum on infinite-length cylinders at a partition into two semi-infinite halves. The MPS ansatz allows for the efficient computation of the ground state Schmidt decomposition:
\begin{align}
    |\Psi\rangle = \sum_\alpha \lambda_\alpha |\Phi_{\alpha L}\rangle |\Phi_{\alpha R}\rangle,
\end{align}
where $|\Phi_{\alpha L/R}\rangle$ are orthonormal bases for the left/right Hilbert spaces. The Schmidt values are non-negative and normalized as $\sum_\alpha \lambda_\alpha^2 = 1,$ with $\lambda_\alpha^2$ being the probability of measuring subsystem $R$ in the state $|\Phi_{\alpha R}\rangle$ (which collapses $L$ to $|\Phi_{\alpha L}\rangle$). Tracing out $R$, one obtains a mixed state for $L$ with von Neumann entanglement entropy
\begin{align}
    S = - \sum_\alpha \lambda_\alpha^2 \log \lambda_\alpha^2.
\end{align}
In Ref.~\cite{LiHaldane2008}, Li and Haldane conjectured that the entanglement spectrum at a bipartition $L\cup R$ of a quantum Hall ground state encodes the low-energy spectrum of the half-system with a physical edge at the location of the bipartition. Given the reduced density matrix $\rho_{L}$, the entanglement spectrum is defined as the eigenvalues of the non-negative-definite Hermitian matrix $H_{L}$ with $\rho_{L}=e^{-H_{L}}$, namely $\epsilon_\alpha = -\log\lambda_\alpha^2$. The Li-Haldane proposal has been further investigated in QH systems~\cite{Thomale2010,Chandran2011,Sterdyniak2012}, demonstrated exactly for free fermions~\cite{Fidkowski2010}, and justified more generally using CFT~\cite{Qi2012,SwingleSenthil2012,Tu2013}.

The entanglement eigenstates can be labeled by their charge, spin, and circumferential momentum quantum numbers. This enables us to define the Schmidt weight in sector $q=(Q,S_z,k_y)$ as the restricted sum $W_q = \sum_{\alpha|q} \lambda_\alpha^2$ over states $\alpha$ with quantum number $q$, so that $\sum_q W_q = 1$. In the infinite geometry, relative charges are unambiguous while the absolute charge is subtle~\cite{Zaletel2013}. For concreteness, we use the following convention: we fix the origin $(Q,S_z)=(0,0)$ so that $Q=0$ has the largest Schmidt weight and the average of $S_z$ is zero over the distribution. In each fixed $Q$ sector, we further place the lowest-lying entanglement eigenvalue at $k_y=0$ for our plots.

In Fig.~\ref{fig:YC6_ES}(a,c) we plot the Schmidt weight resolved by $Q$ and $S_z$ for the $L_y=6$ cylinder ground states at $U/t=6,14$. We perform the bipartition to the right of ring $x=0$. It is found to be strongly peaked at a single $Q$ for the $U/t=14$ state, but not as a function of $S_{z}$. This indicates weak charge fluctuations (i.e., Mott insulator) and relatively large spin fluctuations. Fig.~\ref{fig:YC6_ES}(b), we fix the two most prominent values of $Q$ at $U/t=6$ and plot the ES as a function of $k_y$; these spectra carry integer vs. half-integer spin representations, consistent with being related by a unit charge (\textit{i.e.}, an electron). Fig.~\ref{fig:YC6_ES}(d) shows the ES for the $U/t=14$ CSL state at $Q=0$, the dominant charge sector. An equivalent ES is obtained at every other cut along the cylinder, indicating the lack of translation SSB.

\begin{figure}[h]
    \centering
    \includegraphics[width = \textwidth]{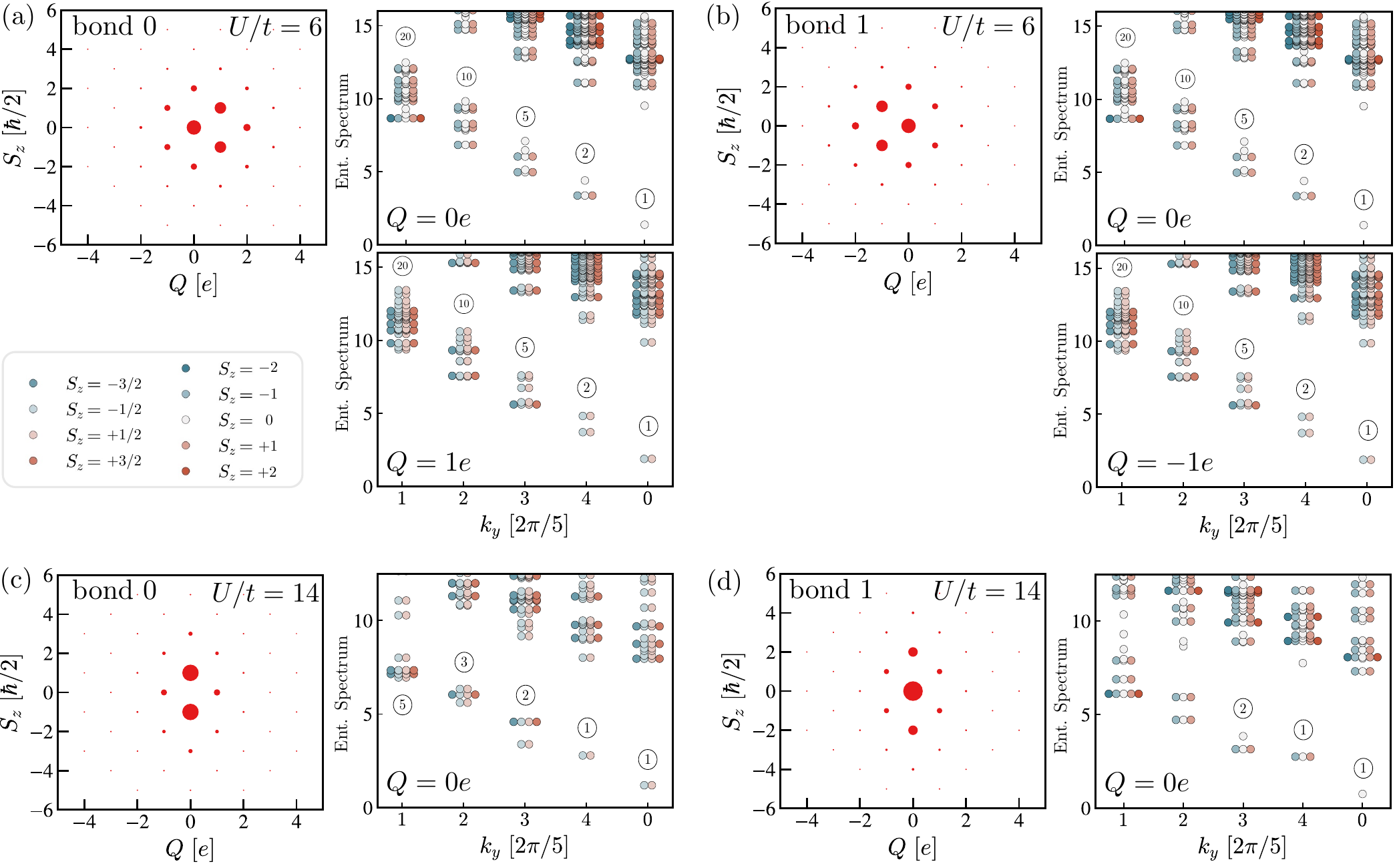}
    \caption{Entanglement data at $L_y=5$ with $\Phi_\text{ext}=0$, analogous to Fig.~\ref{fig:YC6_ES}. (a) For the cut to the right of ring $x=0$ (\textit{i.e.}, on bond 0), at $U/t=6$. Entanglement eigenvalues shown in both the $Q=0e$ (upper right panel) and $Q=+1e$ (lower right) sectors. (b) Likewise at bond 1; entanglement spectrum shown for $Q=0e$ and $Q=-1e$ sectors. (c,d) Similarly for $U/t=14.$}
    \label{fig:YC5_ES}
\end{figure}

In Fig.~\ref{fig:YC5_ES} we plot analogous data for the $L_y=5$ cylinder with $\Phi_\text{ext}=0$, which has the glide-particle-hole symmetry. Within the IQH phase at $U/t=6$, the distribution of Schmidt weights is reflected as a function of $Q$, but nonetheless equivalent, between bond 0 and 1. In contrast, the ES in the $U/t=14$ CSL state differs dramatically at bond 0 vs. 1, with the lowest-lying states carrying half-integer vs. integer spin, indicating SSB of $\mcP T_x$.

\section{Operators in the hybrid $x\, k_y$ representation}

In our analysis of the transition, it is useful to study nearest-neighbour hoppings and their correlations. Since the states of interest are translation-invariant, namely around the cylinder circumference, we are motivated to study the \textit{ring-averaged} correlation functions. For each nearest-neighbour lattice displacement
\begin{align}
    \bm{\delta} \in \{\pm \bm{a}_1, \pm \bm{a}_2, \pm(\bm{a}_1 - \bm{a}_2)\},
\end{align}
let us define
\begin{align}
d^\dag_{\bm\delta}(x) &= \frac{1}{L_y} \sum_y c^\dag(x,y) c((x,y) + \bm\delta).
\end{align}
By Eq.~\eqref{eq:hybridFourier}, 
\begin{align}
d^\dag_{\bm\delta}(x) &= \frac{1}{L_y^2} \sum_y \sum_{k,k'} e^{iky} e^{-ik'(y+\delta_y)} c^\dag(x,k) c(x+\delta_x,k') \\ 
&= \frac{1}{L_y^2} \sum_y \sum_{k,k'} e^{i(k-k')y}e^{-ik'\delta_y} c^\dag(x,k) c(x+\delta_x,k') \\
&= \frac{1}{L_y} \sum_{k} e^{-ik\delta_y} c^\dag(x,k) c(x+\delta_x,k),
\end{align}
so that
\begin{align}
d^\dag_{\bm a_1}(x) &= \frac{1}{L_y} \sum_{k} c^\dag(x,k) c(x+1,k) \\
d^\dag_{-\bm a_1}(x) &= \frac{1}{L_y} \sum_{k} c^\dag(x,k) c(x-1,k) \\ 
d^\dag_{\bm a_2}(x) &= \frac{1}{L_y} \sum_{k} e^{-ik} c^\dag(x,k) c(x,k) \\ 
d^\dag_{-\bm a_2}(x) &= \frac{1}{L_y} \sum_{k} e^{+ik} c^\dag(x,k) c(x,k),
\end{align}
and similarly for $\bm\delta = \pm(\bm{a}_1 - \bm{a}_2)$.

\subsection{Staggered Heisenberg Order Parameter}

The following documents the numerical computation of the staggered Heisenberg order parameter in the hybrid $xk_y$ fermion representation. The spin correlation operator \(\vec{S}_i \cdot \vec{S}_j\) can be expressed in terms of fermionic operators by first noting that 
\begin{align}
    S_i^+ = c_{i\uparrow}^\dagger c_{i\downarrow}, \quad S_i^- = c_{i\downarrow}^\dagger c_{i\uparrow}, \quad S_i^z = \frac{1}{2}(c_{i\uparrow}^\dagger c_{i\uparrow} - c_{i\downarrow}^\dagger c_{i\downarrow}).
\end{align}
The spin correlation operator \(\vec{S}_i \cdot \vec{S}_j\) can then be written in terms of these fermion operators as:
\begin{align}
    \vec{S}_i \cdot \vec{S}_j = \frac{1}{2}\left( c_{i\uparrow}^\dagger c_{i\downarrow} c_{j\downarrow}^\dagger c_{j\uparrow} + c_{i\downarrow}^\dagger c_{i\uparrow} c_{j\uparrow}^\dagger c_{j\downarrow}\right) + S_i^z S_j^z.
\end{align}
Next, we express the staggered-Heisenberg order parameter in the hybrid $(x,k_y)$ operator representation, where $x$ indexes cylinder rings. Because of translation invariance around the cylinder, it is useful to define the $y$-averaged inter-ring NN Heisenberg operator
\begin{align}
\mathcal{O}_{\text{H}}(x)=\frac{1}{L_{y}}\sum_{y}\sum_{a=x,y,z}S^{a}(x,y)S^{a}(x+1,y),
\end{align}
where
\begin{align}
S^{a}(x,y)=c^{\dagger}(x,y)\frac{\sigma^{a}}{2}c(x,y)=\frac{1}{L_{y}}\sum_{k,k'}e^{i(k-k')y}c^{\dagger}(x,k)\frac{\sigma^{a}}{2}c(x,k').
\end{align}
In general, the hybrid representation reads
\begin{align*}
\mathcal{O}_{\text{H}}(x) & =\frac{1}{L_{y}}\sum_{y}\sum_{a=x,y,z}S^{a}(x,y)S^{a}(x+1,y)\\
 & =\frac{1}{L_{y}^{2}}\sum_{a=x,y,z}\sum_{k,k'}\sum_{p,p'}\left(\frac{1}{L_{y}}\sum_{y}e^{i(k-k'+p-p')y}\right) \\
 &\qquad\quad \times c^{\dagger}(x,k)\frac{\sigma^{a}}{2}c(x,k')\cdot c^{\dagger}(x+1,p)\frac{\sigma^{a}}{2}c(x+1,p')\\
 & =\frac{1}{L_{y}^{2}}\sum_{a=x,y,z}\sum_{k,k'}\sum_{p,p'}c^{\dagger}(x,k)\frac{\sigma^{a}}{2}c(x,k')\cdot c^{\dagger}(x+1,p)\frac{\sigma^{a}}{2}c(x+1,k-k'+p).
\end{align*}

\begin{figure}
    \centering
    \includegraphics[width = 494 pt]{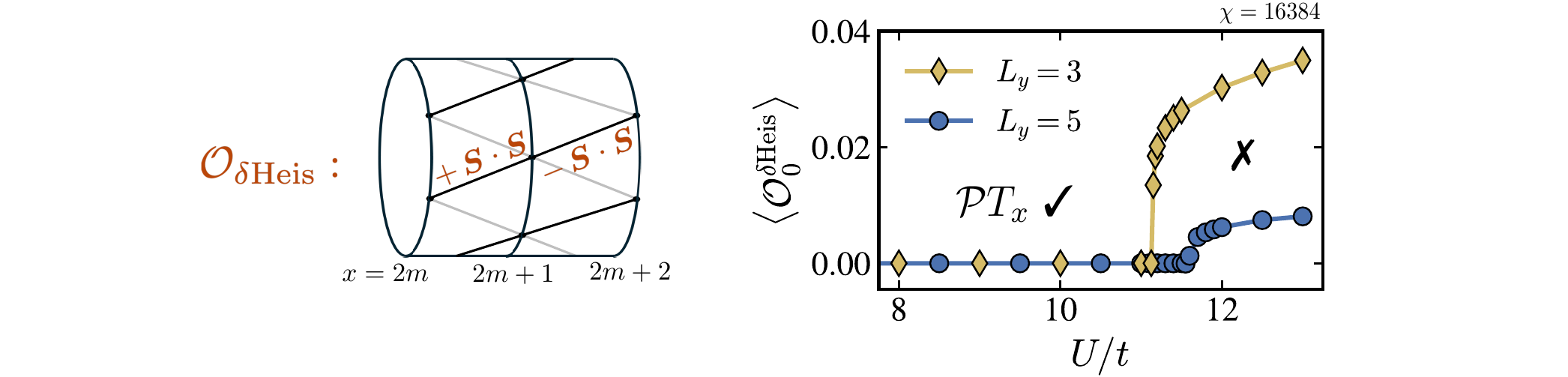}
    \caption{Left panel: Sketch of three rings of cylinder at indicated $x$ values (cylinder ring indices), with solid and dashed lines showing along which staggered-Heisenberg order parameter [defined in main text] changes under glide particle-hole symmetry $P T_x$ on odd-circumference cylinders. Right panel: Magnitude of the staggered Heisenberg order parameter of Eq.~\eqref{eq:dHeis0} plotted against $U/t$ for cylinder widths indicated and with $PT_x$ preserving and broken phases indicated by the check and cross symbols. Consistent with the recovery of this symmetry at the putative 2+1D critical point, the magnitude of the order parameter [bottom left] decreases substantially from $L_y = 3$ to 5.}
    \label{fig:SM_staggered_Heis_OP}
\end{figure}

Since both the IQH and CSL ground states exhibit a two-ring translation invariance (\textit{i.e.}, are invariant under $(T_x)^2$ in addition to $T_y$) then we can measure the following operator defined on the first two bonds near the origin:
\begin{align} \label{eq:dHeis0}
\mathcal{O}^{\delta\text{Heis}}_0 = \mathcal{O}_{\text{H}}(0)-\mathcal{O}_{\text{H}}(1).
\end{align}
In Fig.~\ref{fig:SM_staggered_Heis_OP}, we plot this Heisenberg order parameter on the $L_y=3,5$ geometries with $\Phi_\text{ext}=0$, which have the glide-particle-hole symmetry. This is an alternative to the order parameter computed and displayed in in Fig.~3(a) of the main text, and defined in Eq.~(3), which is instead based on the circumferential current operator. Both order parameters demonstrate that $\mathcal{P} T_x$ symmetry is spontaneously broken in the CSL phase on these two geometries.

\section{Even-circumference supplementary data}
\label{sec:even-Ly-supp-data}

\subsection{Reproducing correlation lengths of Kuhlenkamp \textit{et al.}}
\label{subsec:reproducing-clemens}

\begin{figure}[h]
    \centering
    \includegraphics[width = 494 pt]{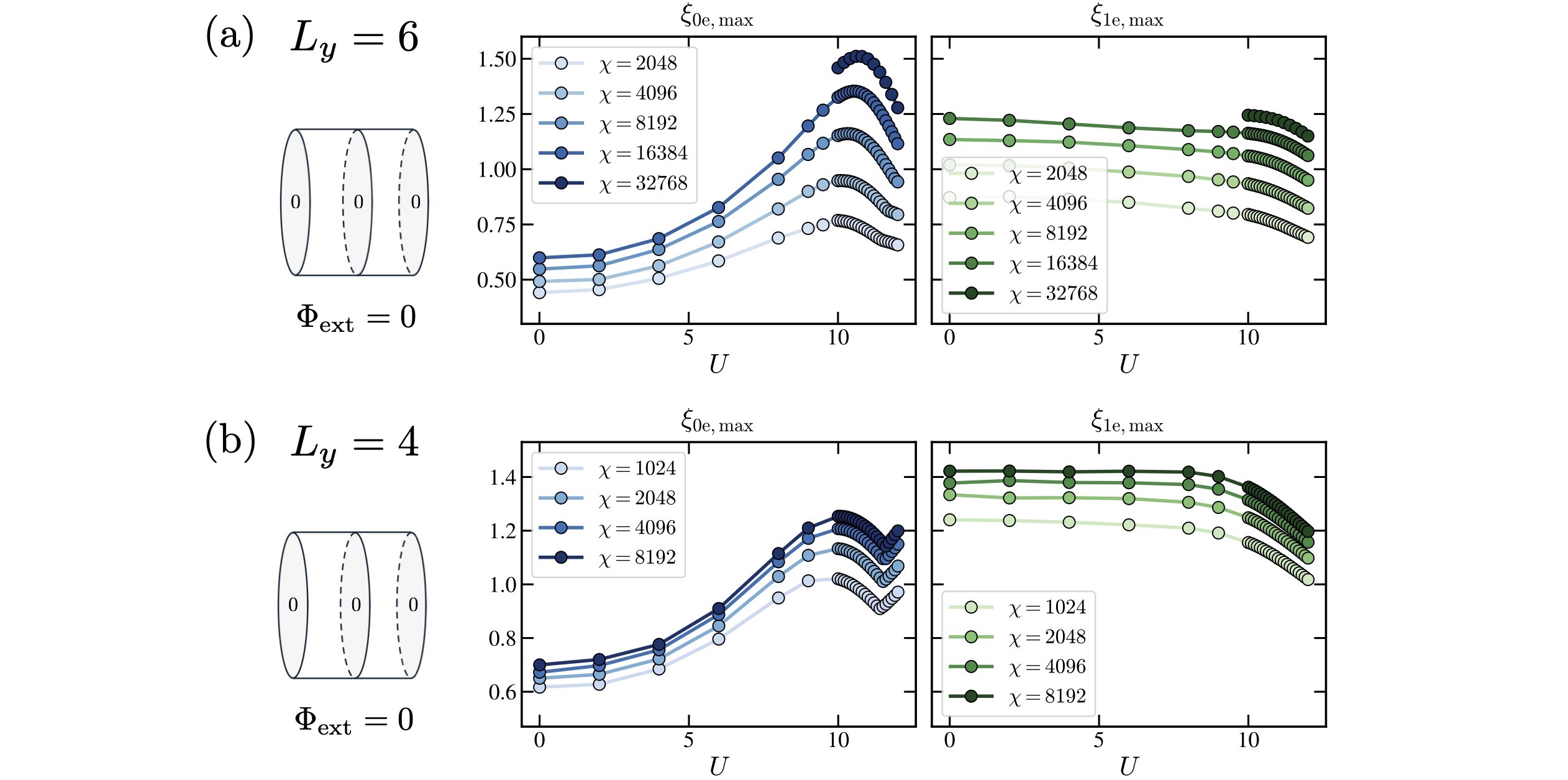}
    \caption{(a) Plot of the maximum correlation length in the charge-$0e$ [middle panel] and charge-$1e$ [right panel] sectors of the ground state transfer matrix for the YC6 geometry (at the particle-hole-symmetric value $\Phi_\text{ext}=0$), with bond dimensions increasing light-to-dark [see legend]. Left illustration: depiction of the flux through the various cylinder rings, consistent with no externally-threaded magnetic flux. (b) Same for YC4, likewise at $\Phi_\text{ext} = 0$.}
    \label{fig:SM_evenLy_xis}
\end{figure}

Here we briefly comment on additional data concerning the correlation lengths for geometries not already highlighted in the main text. In particular, Fig.~\ref{fig:SM_evenLy_xis} shows correlation length data for the $L_y=4$ and $6$ cylinders which are not threaded by external flux.

In particular, $L_y=6$ reproduces the behavior presented in Kuhlenkamp \textit{et al.},~\cite{Kuhlenkamp2024}, pushed here to bond dimension $\chi=32768$. We remark that the correlation lengths are smaller than those with $\Phi_\text{ext} = \pi/2$ external flux exhibited in the main text, but similarly exhibit a maximum in the correlation length in the $(Q,S_z)=(0e,0\hbar)$ sector. The peak (whose height is still increasing with bond dimension) is located at $U/t = 10.6$ at the largest bond dimension $\chi = 32768$; the $U$ location of the peak is also still slowly growing with bond dimension.

On the $L_y=4$ cylinder, the correlation length in the $(Q,S_z)=(0e,0\hbar)$ sector similarly exhibits a finite, local maximum. It is peaked at the smaller value $U/t = 10.0$. There is an uptick in the correlation length at $U/t = 11.7$, which is accompanied by an identical value in the $(Q,S_z) = (0e, \pm \hbar)$ sectors; this feature indicates the softness of a spin-triplet excitation at $L_y=4$, and is sub-dominant in comparison at $L_y=6$. We note that the $Q = 1e$ correlation length does not show any feature in the vicinity of this peak or at the uptick.

\subsection{Flux-threaded system: correlation length extrapolation for main text} \label{sec:xi_extrap_method}

\begin{figure}[h]
    \centering
    \includegraphics[width = 494 pt]{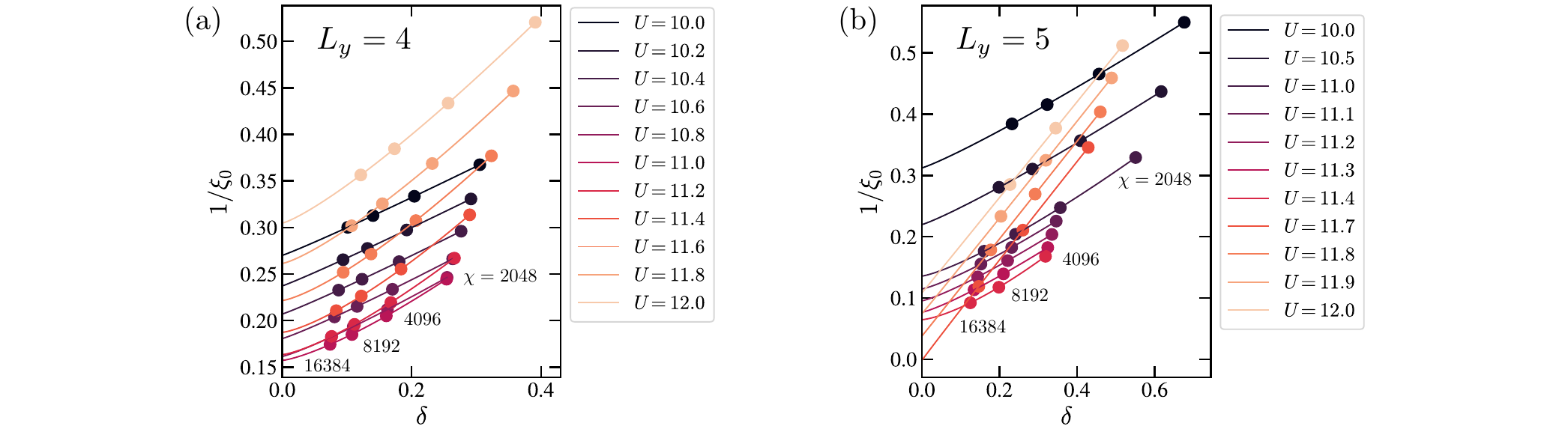}
    \caption{(a) Inverse correlation length in the charge-neutral sector at $L_y=4$ and $\Phi_\text{ext}=\pi/2$. The $y$-intercept of the fitted curve is the correlation length extrapolated to infinite bond dimension; the curve is obtained following the method in Ref.~\cite{Rams2018}, as detailed in Sec.~\ref{sec:xi_extrap_method}, for a range of Hubbard strengths [different colors], sampled from a sequence of increasing bond dimensions [same-colored points moving sequentially to the left]. The extrapolated inverse correlation length saturates at a minimum value exceeding $0.15$, \textit{i.e.}, the correlation length is bounded above by $6.6$ ring spacings.
    (b) The same for $L_y=5$ and $\Phi_\text{ext}=0$.
    }
    \label{fig:SM_xi_extrap}
\end{figure}

To obtain the extrapolated correlation lengths shown in the main text, we employ a recently-developed method for extrapolating correlation lengths in DMRG \cite{Rams2018,Vanhecke2019}.
Namely, we compute several of the largest eigenvalues of the transfer matrix $T X_i = \lambda_i X_i$ in a given charge (\textit{i.e.}, electric charge and spin) sector for each $k_y$, and select those from the same ``excitation branch'' \cite{Rams2018}, namely carrying a consistent complex phase.
These are related to the correlation lengths by
\begin{align}
    \epsilon_i = 1/\xi_i = -\log \n{\lambda_i},
\end{align}
with $\epsilon_i \le \epsilon_{i+1}$. Throughout this work, we report $\xi$ in units of cylinder rings, or equivalently $u = a\sqrt{3}/2$ where $a$ is the lattice spacing of the underlying triangular lattice. Correspondingly, $\epsilon$ is reported in units of $u^{-1}$.
For a system that is gapped in a charge sector $q=(Q,S_z,k_y)$, both $\epsilon_{1,q}, \epsilon_{2,q}$ converge to the same finite value as $\chi\to\infty$, namely the finite inverse correlation length. As a result, the quantity
\begin{equation}
    \delta_q = \epsilon_{2,q} - \epsilon_{1,q}
    \label{eq:scaling_delta}
\end{equation}
vanishes at $\chi\to \infty$, serving as a good ``scaling variable'' that measures the deviation from convergence.
To estimate $\xi(\chi \to \infty)$, we therefore compute $\epsilon(\delta)$ at each $\chi$ (which gives a data point $(\epsilon(\chi),\delta(\chi))$) and extrapolate to $\delta = 0$ by fitting to the form
\begin{align}
    \epsilon(\delta) = a + b\delta^c.
\end{align}
This extrapolation is shown in Fig.~\ref{fig:SM_xi_extrap} for $L_y=4$ at $\Phi_\mathrm{ext}=\pi/2$, yielding the extrapolated values shown in the main text. The correlation length is computed in the $(Q,S_z,k_y)=0$ charge sector, which hosts largest correlation length. We also perform the extrapolation for $L_y=5$ at the glide-particle-hole symmetric flux $\Phi_\mathrm{ext}=0$, which instead exhibits a divergence in $\xi_0$ at a value of $U$ in the range $U\in (11.4,11.8)$; In the main text we take $U_c(L_y=6) = 11.6$, at which the correlation length is largest at the largest bond dimension $\chi = 16384$.

\section{Supplementary odd-circumference analysis}
\label{sec:odd-Ly-supp-data}

\subsection{YC3 Correlation Lengths}
\label{subsec:odd-Ly-corr-length}

\begin{figure}[h]
    \centering
    \includegraphics[width = 494 pt]{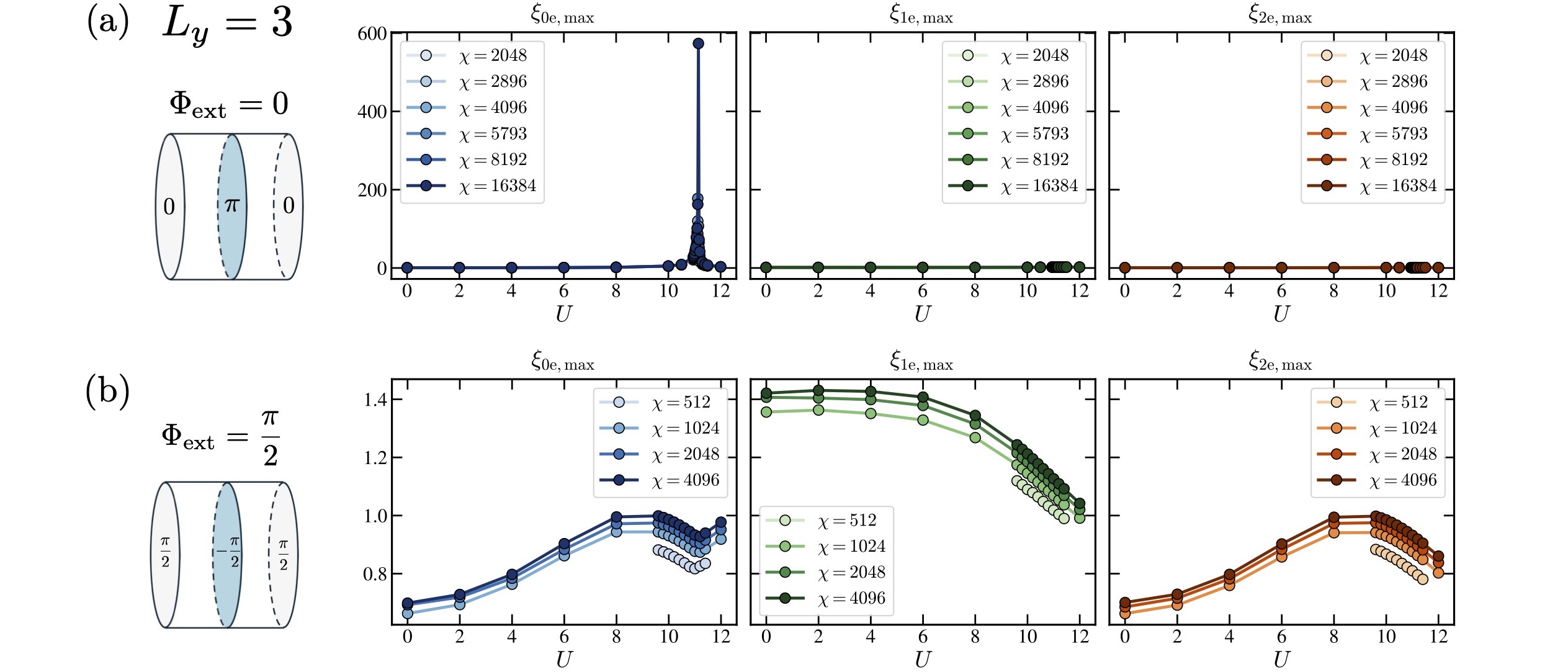}
    \caption{(a) Plot of the maximum correlation length in the charge-$0e$, charge-$1e$, and charge-$2e$ sectors of the ground state transfer matrix for the YC3 cylinder geometry, with bond dimensions increasing light-to-dark. Left illustration: depiction of the flux through the various cylinder rings, consistent with no externally-threaded magnetic flux, $\Phi_\text{ext}=0$. (b) Same for the case $\Phi_\text{ext}=\pi/2$, for which the Hamiltonian explicitly breaks glide-PH symmetry and therefore does not exhibit SSB criticality.}
    \label{fig:SM_Ly3_data}
\end{figure}

In Fig.~\ref{fig:SM_Ly3_data}, we display the correlation lengths for the YC3 geometry. In particular, in panel (a) we provide data for the $\Phi_\text{ext}=0$ case, which accompanies Fig.~3 of the main text. Here, the system respects glide-PH symmetry and, like the YC5 system with $\Phi_\text{ext}=0$ studied in the main text, exhibits a symmetry-unbroken small-$U$ phase and symmetry-broken larger-$U$ phase. The transition between then is evidently continuous, featuring a diverging correlation length which reaches almost 600 ring spacings at the maximum bond dimension $\chi=16384$ considered. In Fig.~\ref{fig:SM_Ly3_data}(b), in contrast, threading $\Phi_\text{ext}$ explicitly breaks the glide-PH symmetry, which reduces the transition to a crossover with a small and wide peak in the correlation length.

\subsection{Relation to ionic Hubbard chain}
\label{subsec:ionic-hubbard}

Here, we make more precise the relationship between the ``ionic Hubbard'' chain and our model on those YC-$L_y$ cylinders (with odd $L_y$ and $\Phi_\text{ext}=0$) \textit{which possess glide-PH symmetry}. In particular, the former model can be thought of as a truly 1+1D version of the latter, \textit{i.e.}, a bipartite chain as opposed to a quasi-(1+1)D system with the hopping-connectivity of a cylinder. We first compare them at the level of symmetries. The ionic Hubbard Hamiltonian is defined as~\cite{Kampf2003}:
\begin{align}
    H = -t\sum_{i,\sigma}(c^\dagger_{i,\sigma}c_{i+1,\sigma}+h.c.) 
    + U\sum_i n_{i,\uparrow} n_{i,\downarrow} + \frac{\Delta}{2} \sum_{i,\sigma} (-1)^i n_{i,\sigma}.
\end{align}
The Hamiltonian is invariant under translation by two sites, with $\Delta$ being the potential energy difference between the two sublattice sites in the unit cell. Moreover, it is invariant under site-centered inversion (but not bond-centered inversion), time-reversal, and a ``duality'' symmetry that is the combination of particle-hole with a single-site translation~\cite{Verresen2021}. These symmetries of the ionic Hubbard chain correspond to the following symmetries of our model: $(T_x)^2,$ site-centered $C_{2z}$, $M_y \mathcal{T}$, and glide-PH $\mathcal{P}T_x$.

Remarkably, the two models are also similar at the level of their ground states. The small-$U$ ground state phase of the ionic Hubbard chain is a fully-symmetric band insulator. Is it widely believed that the phase which appears next upon increasing $U$ is a ``spontaneously-dimerized'' phase which spontaneously breaks the duality and site-centered inversion symmetries~\cite{Verresen2021}. Moreover, as mentioned in the main text, it is believed that these two states are separated by a 2D Ising critical point. This Ising picture has been substantiated both by weak-coupling bosonization arguments~\cite{FabrizioPRL,FabrizioNPB} and an effective strong-coupling effective spin model formulation~\cite{Tincani2009}. However, at least one careful numerical study of the transition using finite DMRG finds features inconsistent with Ising criticality~\cite{Manmana2004}, though the numerical observations are still consistent with a gap to charged excitations and a gapless neutral sector~\cite{Tsuchiizu2004}. These two phases should be identified with the odd-circumference (symmetry-respecting) IQH and (symmetry-breaking) CSL phases, respectively.

\section{Density-density correlations at all circumferences}
\label{sec:density-density-corr}

\begin{figure}
    \centering
    \includegraphics[width=246pt]{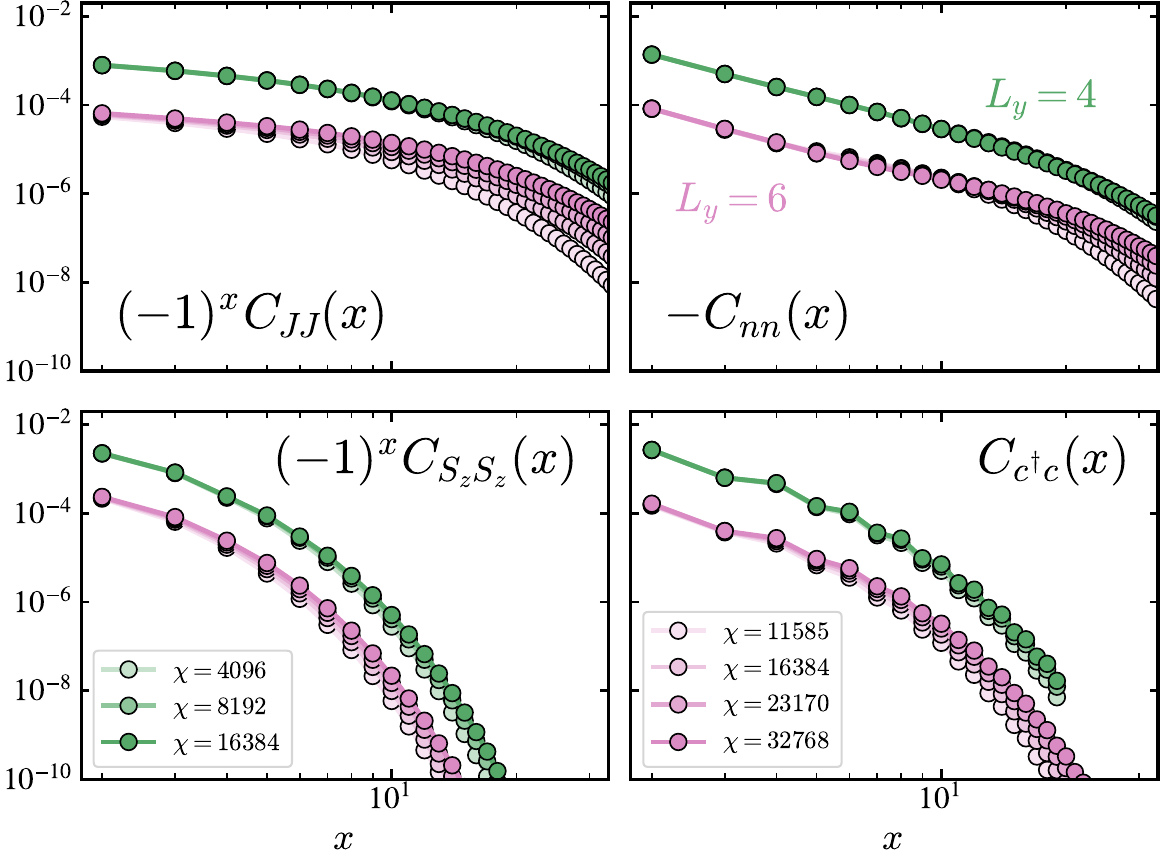}
    \caption{Connected correlation functions for the $L_y=4,6$ systems with $\Phi_\text{ext}=\pi/2$. The Hubbard interaction is tuned to the value where the correlation lengths are maximal, namely $U(L_y=4)=11.0$ and $U(L_y=6)=11.4$. (a) Current-current correlator (defined in main text). (b) Correlator of ring-averaged charge density. (c) Correlator of ring-averaged spin density density. (d) Electron-hole correlator (defined in main text). The $L_y = 6$ correlations are reduced by $\times 10$ for ease of viewing.
    }
    \label{fig:SM_evenLy_corrs}
\end{figure}

In Fig.~\ref{fig:SM_evenLy_corrs} we show various connected correlation functions for the even-circumference cylinders studied in Fig.~3 of the main text, namely $L_y=4,6$ with $\Phi_\text{ext} = \pi/2$.
We plot the current-current (Fig.~\ref{fig:SM_evenLy_corrs}a) and density-density (Fig.~\ref{fig:SM_evenLy_corrs}b) connected correlation functions, which extend far beyond the spin-spin (Fig.~\ref{fig:SM_evenLy_corrs}c) and electron-hole (Fig.~\ref{fig:SM_evenLy_corrs}d) correlators (defined in main text). The first two extend to much longer range than the latter, suggesting both the existence of a spin gap across the transition and the prominent of spin-singlet charge-neutral fluctuations at criticality. While the even-$L_y$ systems always exhibit a crossover from IQH to CSL, their behavior is qualitatively consistent with the odd-$L_y$ cylinders with $\mathcal{P}T_x$ symmetry: the spin-spin and electron-hole correlations are also rapidly exponentially decaying, and the current-current correlator is likewise dominant.

\end{document}